\documentclass[11pt]{article}
\usepackage[margin=0.95in]{geometry}
\usepackage{authblk}
\usepackage{amsmath}
\usepackage{amssymb}
\usepackage{graphicx,setspace,fullpage,fbb,fancyhdr,enumitem,hyperref,slashed,float,physics,mathtools}
\usepackage{tikz} 
\usepackage{cite}
\usetikzlibrary{shapes,arrows,positioning,automata,backgrounds,calc,er,patterns}
\usepackage{xcolor}
\usepackage{subcaption}
\usepackage{tikz-feynman}
\tikzfeynmanset{compat=1.1.0}
\usepackage{adjustbox}
\usepackage{relsize}
\newcommand{\s}{\slashed}

\newcommand{\nh}{\dfrac{\s n}{2}}
\newcommand{\bnh}{\dfrac{\s{\bar{n}}}{2}}

\newcommand{\bbn}{\bar{b}_{n}}
\newcommand{\btn}{\bar{t}_{n}}

\newcommand{\tn}{t_{n}}
\newcommand{\bn}{b_{n}}

\newcommand{\dfun}[1]{\delta\left(#1\right)}
\newcommand{\nbar}{\bar n}

\newcommand{\nn}{\nonumber} 
\newcommand{\beq}{\begin{equation}}
\newcommand{\eeq}{\end{equation}}
\newcommand{\bea}{\begin{align}}
\newcommand{\eea}{\end{align}}
\newcommand{\mo}{\mathcal{O}}
\newcommand{\ml}{\mathcal{L}}
\newcommand{\tor}{\rightarrow}

\begin{document}

\title{{\relsize{-0.5} Resolving the Ultracollinear Paradox with Effective Field Theory}}

\author[1]{Matthew Baumgart}
\author[1]{Panagiotis Christeas}

\affil[1]{\textit{ Department of Physics, Arizona State University, Tempe, AZ 85287, USA}}

\date{}

\maketitle

\abstract{Naive intuition about scale decoupling breaks down in the presence of fermion masses.  Kinematic enhancements can greatly extend the range where one needs to keep a finite mass in calculations to obtain a correct result at even the $\mo(1)$ level.  Treating a light fermion as massive though, leads to a known but somewhat obscure paradox, a seeming leading-order sensitivity to an arbitrarily small mass.  We show how a proper formulation in effective field theory not only resolves the physical conundrum, but answers the very practical question of when fermion masses are required.  This has important implications for the development of shower Monte Carlo above the weak scale.}

%%%%%%%%%%%%%%%%%%%%%%%%%%%%%%%%%%%
\section{Introduction}
\label{sec:intro}
%%%%%%%%%%%%%%%%%%%%%%%%%%%%%%%%%%%

Mathematically, the fact that $\int_0^\infty \frac{m^2}{(k^2 + m^2)^2} dk^2$ = 1 is utterly unremarkable.  The physicist can recognize in this equation though, the seeds of catastrophe, even paradox.  The integrand resembles a fermion propagator squared, with the numerator projected onto the terms from the ``$m$'' piece of $\slashed p + m$.  The integration arises easily enough from loop momenta or phase space.  Standard lore tells us to drop the masses of any fermions that are well below the energy scale of interest, unless one cares about the details of small power corrections.  However, this simple integral reveals a strong discrepancy between zeroing out $m$ before or after the calculation in question.    

This noncommutation of integration and the limit $m \tor 0$ was recognized long ago by Smilga and dubbed a ``quasiparadox'' of massless QED\cite{Smilga:1990uq}.  In fact, the issue plagues general theories of relativistic fermions.  One can push to an absurd limit by imagining a particle with $m = 10^{-10^{10^{10}}}$ GeV.  That would be significantly below current limits on the photon, and yet keeping the $m$ piece of its propagator numerator seemingly leads to an unsuppressed perturbative contribution.  By similar logic, one could imagine using the LHC to determine the Standard Model neutrino masses.

The physical resolution arises from thinking about the production of light fermions at high energies.
In the immediate aftermath of the collision, the fermion is not in an approximately on-shell, asymptotic state (outside of a tiny corner of phase space).
Instead, it begins life as a wavepacket that evolves in time, eventually decaying to the physical fermion plus a cloud of radiation. 
It is these emissions of other particles that introduce terms of the form $\frac{m^2}{(k^2 + m^2)^2}$ . Thus, if we want to experimentally determine the mass
then we must observe this radiation. 
Sensitivity to the enhanced part of the integrand $\frac{m^2}{(k^2 + m^2)^2}$, where the mass effects are important, requires $k_\perp^2 \lesssim m^2$.
As we will detail in calculations below, what enters the denominators in our calculations of interest provides a condition on transverse momentum. 
However, as Smilga points out, generating such ``ultracollinear'' radiation 
takes time, and for a particle with energy $E$ and flight path of length $L$,
there is a lower cutoff on the 
emission angle of $\theta \sim 1/\sqrt{E\, L}$.  We therefore pick up a contribution  
\beq
\int_{k_{\perp {\rm min}}} \frac{m^2}{(k_\perp^2 + m^2)^2} dk_\perp^2 \sim \frac{m^2}{E^2 \theta^2} \sim \frac{m^2 \, L}{E},
\label{eq:length}
\eeq
with $k_{\perp {\rm min}} \sim \sqrt{E/L} \gg m$.  This resolves the strong form of the paradox as sensitivity to particle masses now depends sensibly on the size of the experiment.  Independent of detailed technical considerations, this places a lower limit on fermion mass measurable by the LHC, $m_{\rm LHC,\,min} \sim \mo(100)$ eV. 

We have avoided a deep physical conundrum, and yet the situation remains unsatisfying.  Taken at face value, the result of Eq.~\ref{eq:length} implies that we still must track fermion masses at scales substantially below those of our process's center of mass energy.  For computations at the LHC, one might worry that every neglected charged-lepton mass results in a sizable error.  It is the purpose of the present paper to allay these fears, in general, but also to clarify the energy regime where ultracollinear effects are important, providing a systematic treatment of how to include them. This will show where masses must be kept in a practical computation.

We see from Eq.~\ref{eq:length} that the relevant question concerns the scale of $k_{\perp  {\rm min}}$.  The physical size of the experiment provides {\it an} IR cutoff, but it is not necessarily {\it the} IR cutoff for the problem at hand.  For example, light fermions produced well above the weak scale have the ability to radiate on-shell electroweak bosons.  Thus, in this regime, particle virtualities are regulated in the IR by something at least at the order of $m_W$, which dominates all fermion masses except the top quark.  Below this scale, with weak particles integrated out, but still well above the fermion mass, $m_f$, one must nonetheless account for the vanishingly small phase space to produce nearly on-shell light particles in high-energy processes. 
This is quantified by the ubiquitous Sudakov factors found in observables that determine the (non)showering of radiation.\footnote{ There are subtleties that arise in the electroweak theory, which still demands gauge-invariant external states because of Elitzur's Theorem \cite{Elitzur:1975im}, in spite of spontaneous symmetry breaking.  This can lead to cancellations of Sudakov logs that otherwise naively appear \cite{Maas:2022gdb}.  One may need to carefully construct gauge-invariant electroweak final states as in \cite{Nabeebaccus:2022jhu}. Nonetheless, the running of fragmentation functions and similar operators can generate exponentiated logs in observable calculations \cite{Manohar:2014vxa,Baumgart:2014vma,Ovanesyan:2014fwa,Bauer:2014ula,Baumgart:2017nsr,Beneke:2018ssm,Baumgart:2018yed}.  Furthermore, the present work analyzes a pure Yukawa theory, so the resolution of the ultracollinear paradox in this case cannot arise from a careful imposition of gauge invariance on external states.}

Unlike collisions (or explosions) of macroscopic objects, which tend to spray debris in all directions, the scattering of high-energy particles produces well-collimated jets of radiation.  One finds an explanation for this phenomenon in the structure of Feynman diagrams.  The internal propagators of the showering particles provide kinematic enhancement if they are nearly on-shell, as is typically the case for soft or collinear emissions.  This provides a perturbative origin for jets, but it leaves a further puzzle: {\it why then isn't every emission as collinear as possible?}.  Instead, the process is dominated by ``strong-ordering,'' as implemented in shower Monte Carlo like Pythia by either angular or virtuality ordering, with the widest or most off-shell emissions coming first \cite{Bierlich:2022pfr}.  The differential cross section one gets by demanding the absence of real emissions must still account for kinematically-enhanced virtual corrections to all orders of perturbation theory. This gives the overall Sudakov factor, which schematically modifies a Born-level cross section $d \sigma_0 \tor d \sigma_0 \exp[-C \frac \alpha \pi \log(Q^2/m_{\rm IR}^2)]$ in the toy Yukawa theory we study in this work of interest, where $C$ is an $\mo(1)$ number and $m_{\rm IR}$ is parametrically near the scale of observation, typically a particle mass or kinematic cut.  Gauge theories famously have stronger ``double-log'' soft-collinear enhancements, giving $d \sigma_0 \tor d \sigma_0 \exp[-C \frac \alpha \pi \log(Q^2/m_{\rm IR}^2)^2]$.  These overall terms thus give the ``no-split'' probability, as nicely summarized in \cite{Schwartz:2014sze}.  We thus see that for particles with masses $m_f \ll Q$, the rate to produce them from the high-energy collision exactly on-shell gets highly suppressed.

The Sudakov factors show that even in the absence of unambiguous lower physical scales like masses, the physics of jets dynamically produces a soft scale, the ``jet mass'' ($m_J$), that can be parametrically large compared to $m_f$.  We propose in this work, a threshold of \\ $\exp[-C \frac \alpha \pi \log(Q^2/m_{\rm IR}^2)^a] < 50\%$, where $a = 1,2$ depending on the theory.  If the threshold is crossed for $m_{\rm IR} > m_f$, then the particle is more likely to have split than not, and one then works in an effective field theory (EFT) where the fermion is treated as massless, which we take to be SCET1 for the appropriate particles ({\it cf.}~\cite{Bauer:2000ew,Bauer:2000yr,Bauer:2001ct,Bauer:2002uv} for the foundational papers on SCET, or Soft-Collinear Effective Theory).  Once one gets to a jet scale where $\exp(-C \frac \alpha \pi \log(m_J^2/m_f^2)) > 50\%$, then at $m_J$ one matches to a SCET2 theory where that particular fermion is massive.  Lastly, if one drops below $m_f$, then one integrates out the particle.\footnote{We leave it for future exploration whether data is best explained by some other threshold criteria, like having a no-split probability $< 32\%$, 5 $\%$, {\it etc.}  Regardless of the details, we necessarily have a regime where fermions need to be included as propagating degrees of freedom, but with their finite masses accounted for.}  This approach thus systematically improves the crude approximation of taking a particle to either be massless at high energies or integrated out at low energies.  Letting the Sudakov factor decide when to keep $m_f$ in a calculation can greatly modify the naive estimate that once $m_f^2/Q^2 \ll 1$, we are effectively massless.  In particular, we see that the strength of the interaction must be accounted for.  For example, if it is very feeble, then producing a light particle in a high-energy process can lead directly to on-shell states since the amplitude to radiate anything would then be highly suppressed.

The natural arena for the correct treatment of ultracollinear physics is the regime of high-energy multi-particle emissions in colliders and astroparticle systems.  One process of particular interest to us is the indirect-detection of dark matter.  In these processes, it is possible to create Standard Model particles at enormous energies, after which one observes a single energetic particle like a photon, and is inclusive over the rest of the event. In \cite{Chen:2016wkt}, the authors developed a general-purpose electroweak shower Monte Carlo, emphasizing the role that ultracollinear corrections play in the proper handling of both next-to-leading-log (NLL) resummation and next-to-leading-order (NLO) matching.  They found that the (im)proper treatment of fermion masses affects observables at the few-\% level for $\mo(10 \,\text{--}\, 100)$ TeV colliders, which is well within the scope of current theoretical precision.  Furthermore, the discrepancy can grow with energy scale, meaning that astroparticle studies at even higher energies are especially sensitive. Since the focus of \cite{Chen:2016wkt} is a practical realistic shower algorithm, they did not generalize the conceptual issues and questions that surround ultracollinear enhancement.\footnote{One can find an early discussion of the ultracollinear corrections to splitting functions in \cite{Ciafaloni:2010ti}.  As in \cite{Chen:2016wkt}, they did not explore the conceptual consequences of having finite-mass corrections that do not disappear as $m_f \tor$ 0.} This is our concern.  To disentangle this phenomenon as much as possible from the enormous amount of nontrivial physics captured in a parton shower, we work in a simple, perturbative Yukawa theory. We even find even in this case that finite-mass contributions can lead to large, $\mo(1)$ numerical shifts to observables.   

The calculation of showering radiation is done by Monte Carlo in practice.  It is a useful question for future study to what extent finite mass effects and the ultracollinear regime are receiving proper treatment. It is not the case that practical shower routines have manifestly and universally accounted for this physics. {\it Its correct implementation would certainly be in keeping though, with the present sophistication of modern parton showers}.  For example, recent efforts of the community include further QCD and EW corrections along with their resummation \cite{Bauer:2020jay},\cite{Han:2022laq},\cite{Pagani:2023wgc},\cite{Jezo:2023rht},\cite{Assi:2023rbu}, non-perturbative physics such as hadronization, final state radiation (fragmentation functions) and initial state radiation (parton distribution functions) modeling \cite{Han:2022laq},\cite{Pagani:2023wgc},\cite{Assi:2023rbu}, and taking into account mass effects of particles \cite{Jezo:2023rht},\cite{Assi:2023rbu},\cite{Gavardi:2023aco}. Additionally, by going beyond next-to-leading-log (NLL) resummation in observables dominated by soft emissions, \cite{FerrarioRavasio:2023kyg} fills an important gap in building general NLL parton showers.   

The importance of finite-mass corrections to splitting functions has long been recognized, such as in the  quasi-collinear limit of \cite{Catani:2000ef,Cacciari:2002xb}, which gives corrections that nonetheless disappear in taking $m_f \tor $ 0.  This is distinct from our ultracollinear limit, where the conceptual challenge lies in the fact that subleading terms remain even if one takes the massless limit after obtaining them.  Many modern shower algorithms that include the electroweak sector account for particle masses, as weak-scale and heavy quark masses are phenomenologically relevant.  This can be seen in recent updates to both the Vincia \cite{Kleiss:2020rcg} and Herwig \cite{Bewick:2023tfi} showers.  In addition to the quasi-collinear limit, Vincia also employs quasi-soft corrections, which shift the momenta that appear in eikonal factors by finite mass corrections ({\it cf.}~\cite{Platzer:2022nfu} for a systematic discussion of the quasi-collinear \& quasi-soft limits in electroweak physics). 

%The quasi-collinear limit although it seems equivalently with SCET, as it provides, a ratio between invariant masses as constant parameter, it does not provide a flexible framework of an EFT, that would allow the mass to enter systematically, but it restricts itself in the level of amplitudes. We take advantage of the powerful formalism of SCET reaching the same result. Also, mass corrections from the quasi-collinear limit appear as subleading with respect to the mass, in the case of TMD, while in our case, as we study a more inclusive object, as the FF, we found to be actually significant of Leading order, something that points the novelty of this work.}

%When fragmentation functions are considered, especially in EW theory, the massless case is strictly considered so far \cite{Bauer:2020jay,Han:2022laq}. 

There is also the matter of better precision and higher accuracy in the computation which is related to developing the computational tools and algorithms used for the analysis. More specialized approaches have been examined that include SCET \cite{Gavardi:2023aco}, the complex mass scheme to treat finite width effects\cite{Jezo:2023rht},\cite{Gavardi:2023aco}, or alternative jet techniques \cite{Gavardi:2023aco},\cite{vanBeekveld:2023chs}. Good agreement with experimental results have been achieved so far \cite{Assi:2023rbu},\cite{Gavardi:2023aco}.  {\it A priori}, except for \cite{Ciafaloni:2010ti,Chen:2016wkt}, the explicit discussions of mass in these references are different from our ultracollinear concern.  Along with finite-width effects, we can treat them systematically within EFT, where fragmentation functions enter as particular operators.  

The aim of the present work is to focus on the subtle physics of having $m_f^2/(k_\perp^2 + m_f^2)^2$ contributions to observables, disentangling them from the many other technical complications that arise in properly showering the full Standard Model.  To this end, we work with a ``simplest nontrivial example,'' a Yukawa theory consisting of one massless scalar (``higgs''), one massless fermion (``bottom''), and one fermion (``top'') with mass $M_t$.  We use these particle names hereafter, with the understanding that they are \textit{not} strictly the Standard Model versions. This allows for three different regimes of jet physics which depend on the energy scale relative to $M_t$ ({\it cf.}~Fig.~\ref{fig:schem}).  We will see that properly accounting for the mass in the intermediate-regime theory requires inclusion of the top width, $\Gamma_t$, in its propagator.  This shifts splitting functions at $\mo(\alpha_y)$ and is generally more important numerically and more universal (since this effect occurs for massive particles of any spin) than the ultracollinear terms with which we motivated our study.   However, we continue to emphasize the ultracollinear regime as its treatment is more subtle and forces one into very careful thinking about the physics of massive particles in high-energy processes.  

The structure of the paper is as follows, in Section \ref{sec:frag} we review the calculational setup; Appendix \ref{app:ff} provides further details of the fragmentation formalism, which allows us to  calculate observables in the presence of multiple emissions.  Section \ref{sec:frag} also gives the specific implementation of SCET in our Yukawa theory.  This EFT allows for efficient computation of the fragmentation quantities across a wide variety of scales and the resummation of large logarithms that results.  The treatment of SCET soft modes is discussed in Appendix \ref{app:soft}.  In Section \ref{sec:calc}, we sketch some of the calculational ingredients, leaving details for Appendix \ref{app:details}.  Having obtained our quantities of interest, we combine them at leading-log (LL) and a subset of NLL corrections with NLO matching in Section \ref{sec:rm} and Appendix \ref{app:ad} before presenting numerical results and concluding in Section \ref{sec:res}.

%%%%%%%%%%%%%%%%%%%%%%%%%%%
\section{A Toy Theory for Ultracollinear Physics}
\label{sec:frag}
%%%%%%%%%%%%%%%%%%%%%%%%%%%

We begin by specifying our observable of interest.  Thanks to factorization, we are able to write our normalized differential decay rate as a convolution of two separate functions, the Wilson Coefficient $C_i$ and the Fragmentation Function $D_{ij}$ where we capture the high-, and low-energy physics, respectively,
\begin{align}
	\dfrac{1}{\Gamma_0}\dfrac{d\Gamma_{j+X}}{dz} = \int_z^1 \dfrac{dx}{x} C_i \left(x/z\right) D_{ij}\left(z\right).
\label{eq:obsdef}
\end{align}
We are thus interested in the semi-inclusive differential rate to observe particle $j$, carrying a fraction, $z$, of its mother's plus-momentum, where $p^+ = E + p^z$.  Though we have written it as the decay rate of some heavy particle, it is trivial to replace $\Gamma \tor \sigma$ and discuss instead a normalized differential cross section. The normalization factor $\Gamma_0$ is typically determined from the total decay rate for a process of interest or sometimes its tree-level value.  It needs specification for a particular calculation.  In our case, $\Gamma_0$ is the all-orders two-body decay rate of the heavy particle. 

In Appendix \ref{app:ff}, we provide a brief recap of the fragmentation formalism and its implementation in SCET.  In particular, the $D_{ij}$ fragmentation functions  are computed as operator matrix elements.  When subject to only leading-log running in the massless case, they give the probability for $i$ partons to split into $j$ partons.  However, we will see that this interpretation breaks down at higher perturbative order and in the presence of finite particle width.  By matching our calculation at some scale, we determine the Wilson coefficients, $C_i$ for each species of parton.  This gives a boundary condition for renormalization group (RG) evolution.  The usual invariance of the observable, LHS, with change in renormalization scale gives us equations for $C_i$ or $D_{ij}$ that we can use to run to other scales, resumming large logarithms in the process.  This is further detailed in Section \ref{sec:rm}.

%%%%%%%%%%%%%%%%%%%%%%%%%%%
%\section{Effective Field Theory}
%\label{sec:scet}
%%%%%%%%%%%%%%%%%%%%%%%%%%%

\begin{figure}[H]
\centering
\includegraphics[width=0.7\textwidth]{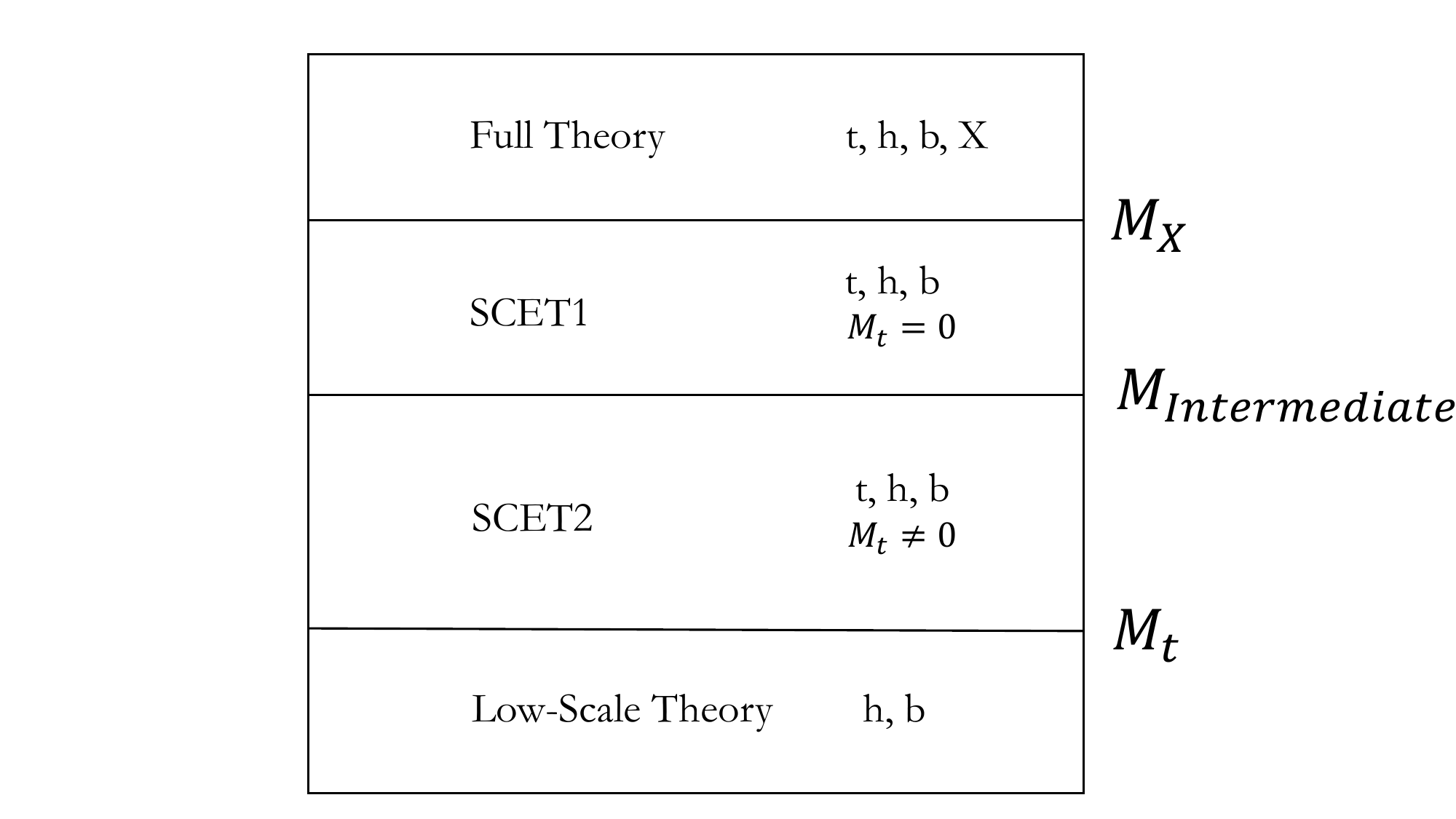}
\caption{Schematic description of our sequence of theories}
\label{fig:schem}
\end{figure}

Our toy theory consists of three particles inspired by their Standard Model namesakes: 1) one massive Dirac fermion, $top$, that is in a doublet of an SU(2) global symmetry; 2) one massless SU(2)-singlet Dirac fermion, $bottom$; 3) one masslesss SU(2)-doublet complex scalar, $higgs$. They have  one flavor off-diagonal Yukawa interaction,
\begin{align}
 \ml = \bar{t}\left( i\slashed \partial - M_t \right)t + \bar{b}\left(i\slashed \partial\right)b +(\partial^\mu h)^\dagger(\partial_\mu h) + y_t\, h \, \bar{t}\, b + h.c.
\label{eq:toyl}
\end{align}
To initiate the fragmentation process, we produce tops at high-energy. Though the fragmentation functions and splitting functions that evolve them are agnostic to the production details, for concreteness, we assume a scalar, $\chi$, with $M_\chi \gg M_t$ that decays exclusively to $t \bar t$ via,
\beq
\ml_{\chi} = y_\chi\, \chi \, \bar t \,t.
\label{eq:xl}
\eeq
At a qualitative level, because the $\chi$ decays into a significantly lighter sector, what it ultimately produces are jets of $t,\,b\, h$ particles.  To accurately describe their contents, we work in four different energy regimes ({\it cf.}~Figure \ref{fig:schem}): 1) A full theory consisting of eqs.~\ref{eq:toyl} \& \ref{eq:xl}, along with $\chi$ kinetic terms; matching to this determines the high-energy boundary conditions for $C_{t,b,h}$. 2) Integrating out the $\chi$, the kinematics of its top-particle decay products at very high energies is dominated by collinear splittings.  We can easily resum the resulting large logs by working in SCET1 with all massless particles. 3)  As the jets evolve and particle fragmentation decreases energies, we enter a regime where we can no longer neglect $M_t$.  This is described by SCET2 with a soft scale $\sim M_t$.  4) Lastly, passing below $M_t$, we integrate out the top, leaving a theory of $b,\,h$ particles that only interact via irrelevant operators.\\
Our interest is only the collinear sectors of SCET1 and SCET2.  We write here the SCET2 collinear lagrangian, from which one can easily get the SCET1 collinear sector by setting $M_t = 0$.
\begin{align}
 \ml_{\text{SCET}}&=  \btn\left\lbrace i \,(n \cdot \partial) +   i \slashed \partial_\perp \dfrac{1}{\nbar \cdot \mathcal{P}} i \slashed \partial_\perp \right\rbrace\bnh\tn - M_t^2 \, \btn \dfrac{1}{\nbar \cdot \mathcal{P}} \tn +\nn \\
 &+ \bbn\left\lbrace i \,(n \cdot \partial) +   i \slashed \partial_\perp \dfrac{1}{\nbar \cdot \mathcal{P}} i \slashed \partial_\perp\right\rbrace\bnh\bn + (\partial_\mu h_n)^\dagger(\partial^\mu h_n) \nn \\
& + y \, \btn \left\lbrace (i \slashed \partial_\perp - M_t)\dfrac{1}{\nbar \cdot \mathcal{P}} h_n - h_n \dfrac{1}{\nbar \cdot \mathcal{P}} i \slashed \partial_\perp \right\rbrace \bnh\bn  \nn \\
& + y \, \bbn \left\lbrace i \slashed \partial_\perp \dfrac{1}{\nbar \cdot \mathcal{P}} h^\dagger_n - h^\dagger_n \dfrac{1}{\nbar \cdot \mathcal{P}} ( i \slashed \partial_\perp +M_t ) \right\rbrace \bnh\tn  \nn \\
 &- y^2 \left\lbrace h_n \btn \dfrac{1}{\nbar \cdot \mathcal{P}}  \bnh \tn h^\dagger_n + h^\dagger_n \bbn  \dfrac{1}{\nbar \cdot \mathcal{P}} \bnh \bn h_n  \right\rbrace.
 \label{eq:scetl}
\end{align}
By the usual convention the label momentum operator $\nbar \cdot \mathcal{P}$ only acts on fields to its right.  Also by standard SCET practice, the two lightcone vectors $n^\mu,\bar{n}^\mu$ are defined as $n^2 = \bar{n}^2 =0$ and $n\cdot\bar{n} = 2$. They form a lightcone basis where we can express any (momentum) vector as $p^\mu = (\bar{n} \cdot p, n \cdot p, p_\perp)= (p^+,p^-,p_\perp),$ and for collinear momenta in SCET1 or 2, $(p^+,p^-,p_\perp) \sim (1,\lambda^2,\lambda)Q$, where $Q$ is a hard scale in the problem of interest, and $\lambda$ is the usual SCET power-counting parameter. Collinear fields in either theory scale with $t_n,b_n,h_n \sim \lambda Q$.  In the SCET2 theory, we also take $M_t \sim \lambda Q$, and thus it must be systematically kept in calculations.  By contrast, in SCET1, $M_t \ll \lambda Q$, and we drop it as it is parametrically below the IR scale of that EFT.  In SCET1, soft momenta $(p^+,p^-,p_\perp) \sim (\lambda^2,\lambda^2,\lambda^2)Q$, while in SCET2 $(p^+,p^-,p_\perp) \sim (\lambda,\lambda,\lambda)Q$. In SCET1, soft fermion fields scale like $t_s,b_s \sim \lambda^3 Q$, while in SCET2, $t_s,b_s \sim \lambda^{3/2} Q$.  For the soft Higgs, in SCET1, $h_s \sim \lambda^2 Q$, and in SCET2, $h_s \sim \lambda Q$.  The lagrangian in Eq.~\ref{eq:scetl} gives the following Feynman rules:

\[
\begin{minipage}{.5\linewidth}
  \centering
\begin{align*}
\begin{tikzpicture}[baseline=-0.35 ex]
\begin{feynman}
\vertex (a);
\vertex [right=of a] (b);
\vertex [above right=of b] (d);
\vertex [ below right=of b] (e);
\diagram* {
(a) -- [double,double distance=0.5ex,thick,with arrow=0.5,arrow size=0.15em,momentum=\(p\)](b),
(b)-- [fermion,rmomentum=\(k\)](d),
(b)-- [scalar,rmomentum'=\(q\)](e),
};
\end{feynman}
\end{tikzpicture}= (+iy)\left[\left(\dfrac{\s k_\perp}{k^+}-\dfrac{\s p_\perp}{p^+}\right) - \dfrac{M_t}{p^+}\right]\bnh
\end{align*}
\begin{align*}
\begin{tikzpicture}[baseline=-0.35 ex]
\begin{feynman}
\vertex (a);
\vertex [above left=of a] (b);
\vertex [below  left=of a] (c);
\vertex [above right=of a] (d);
\vertex [ below right=of a] (e);
\diagram* {
(b) -- [double,double distance=0.5ex,thick,with arrow=0.5,arrow size=0.15em,momentum=\(p\)](a),
(c)-- [scalar,momentum=\(q\)](a),
(a)-- [double,double distance=0.5ex,thick,with arrow=0.5,arrow size=0.15em,rmomentum'=\(l\)](d),
(a)-- [scalar,rmomentum'=\(k\)](e),
};
\end{feynman}
\end{tikzpicture}= (-iy^2) \left(\dfrac{1}{p^+ + k^+}\right)\bnh.
\end{align*}
\end{minipage}
\]
The double-line indicates a $t$, the single line a $b$, and the dashed line an $h$.  One should note that these Feynman rules assume all incoming momenta.  In practice, we will only need the three-point vertex above.  To the order we are working, the four-point vertex\footnote{There is also another four-point vertex that is the same as the one written above, where the $t$ is substituted by $b$.} could only contribute in a virtual process, but as we show in Section \ref{subsec:otlffc}, it does not affect the fragmentation functions.  

One might also worry about the need for soft interactions.  However, with the field and momentum power counting given above, the only leading-power Yukawa vertex that conserves momentum is in SCET1 with collinear $t$ and $b$ fields and a soft $h$.  As we demonstrate in Appendix \ref{app:soft}, in SCET1, where we can neglect $M_t$, the soft limit of the collinear interactions given in Eq.~\ref{eq:scetl} itself reproduces the soft limit of the full-theory contributions to fragmentation amplitudes.  Thus, the contribution arising from the soft vertex is removed by zero-bin subtraction \cite{Manohar:2006nz}.
% In SCET2, the soft-collinear vertex that would contribute is power-suppressed.  
We are therefore left with a straightforward set of all-collinear calculations we detail in Section \ref{sec:calc} and Appendix \ref{app:details}.

%%%%%%%%%%%%%%%%%%%%%%%%%%%
\section{Finite Mass Effects at LO \& Beyond}
\label{sec:calc}
%%%%%%%%%%%%%%%%%%%%%%%%%%%

In Section \ref{sec:frag}, we discussed the fragmentation function, $D_{ij}$, our main calculational quantity of interest.  We provide here an overview of its computation in SCET1/SCET2, and how finite particle masses affect results beyond mere power corrections.  In using the fragmentation formalism to compute particle production at a variety of energy scales, it becomes necessary to run the operators whose matrix elements give $D_{ij}$.  As we will show in Section \ref{sec:rm}, the anomalous dimension, $\gamma_{ij}$, for $D_{ij}$ is
\begin{equation}
\gamma_{ij} = -\dfrac{dD_{ik}}{d\log(\mu)} \, (D)^{-1}_{kj}.
\label{eq:gammadef}
\end{equation}
Fragmentation Functions are calculated through a fixed order diagram based on the definitions in Eq.~\ref{eq:momff}. For example, we can take the $D_{th}$ computation at $\mathcal{O}(\alpha_y)$:
\vspace{0.2in}
\begin{align}
D_{th} = \begin{tikzpicture}[baseline=-0.35 ex]
\begin{feynman}
\vertex (a)[crossed dot] {};
\vertex [right=of a] (b);
\vertex [above right=of b] (c1);
\vertex [ right=of c1] (c2);
\vertex [below right=of c2] (d);
\vertex [ right=of d] (e)[crossed dot]{};
\diagram* {
(a) -- [double,double distance=0.5ex,thick,with arrow=0.5,arrow size=0.15em,momentum=\(p+k\)](b) -- [scalar, momentum=\(p\)](c1),
(c2) -- [scalar, momentum=\(p\)](d)--[double,double distance=0.5ex,thick,with arrow=0.5,arrow size=0.15em,momentum=\(p+k\)](e),
(b)-- [fermion,half right, looseness=1,insertion={[size=10pt]0.5},momentum'=\(k\)](d)
};
\end{feynman}
\end{tikzpicture}
\end{align}
By construction, this process consists of an energetic mother parton fragmenting to a daughter.  In the SCET, the mother is created by a collinear field.  We can investigate various contributions for the other states from the soft and collinear sectors by setting their momenta in these regions.  By energy conservation, one or both of the daughter particles is necessarily collinear.  In SCET2, the soft emission cannot conserve minus-momentum, while the SCET1 soft contribution is power-suppressed by the field scaling given in Section \ref{sec:frag}.
Therefore, only the all-collinear process is the leading one of interest. As a consequence of the power counting in SCET1, we drop $M_t$ as a scale below that of the theory's soft scale. To regulate IR divergences, we impose a fixed $\mu_\textrm{IR}$ cutoff.  Our SCET1 splitting functions just reproduce the standard massless splitting functions  \cite{Ciafaloni:2010ti}.

Once the scale is sufficiently close to $M_t$ that on-shell top production is likely, we match to the intermediate-scale theory, SCET2.  We can no longer neglect $M_t$, and its presence modifies our results at LO because of its finite width, and at NLO through the ultracollinear contributions.  When the shower reaches the scale $M_t$, we integrate out the top. In this Low-Energy Theory, there are only trivial fragmentation functions for the remaining particles, $\sim\delta(1-z)$.

%%%%%%%%%%%%%%%%%%%%%%%%%%%
\subsection{Outline of the FF Computation}
\label{subsec:otlffc}
%%%%%%%%%%%%%%%%%%%%%%%%%%%

The computation of a real-emission fragmentation function will be outlined with an example of $D_{th}$ in SCET2.\footnote{One can obtain the analogous SCET1 $D_{th}$ result by setting $M_t=0$ in the SCET2 integrand and imposing a fixed IR cutoff on perp momentum, $\mu_\textrm{IR}$.  For notational simplicity though, we always call the IR scale in the log argument $M_t$.} Starting from the definition of the Fragmentation Function given in Eq.~\ref{eq:momff}, 
\begin{align}
D_{th}(z) &= \frac{z}{4Q} \Tr \int d\Pi_{X} \, \delta \left(1-\frac{p_X^+ + p^+}{Q}\right) \langle 0 | \bnh t(p+p_X) | h(p)\, X \rangle \langle h(p)\, X | \bar t(p+p_X) | 0\rangle \, \Big |_{p_\perp = 0},
\label{eq:dth}
\end{align}
where we have used $N_c = 2$ since the initiating $t$ is a doublet of SU(2).  The $X$ just arises from the semi-inclusive final state in which the $h$ is observed.  We integrate fully over the $X$ particles' phase space subject to the constraints imposed by the $\delta$-function and on-shellness.  At $\mathcal{O}(\alpha_y)$ we need to compute the graph shown at the start of this section. 
As we can see, the graph is cut on the observed daughter and recoiling particle lines, which puts these particles on shell.  The unintegrated momentum $p$ is thus on-shell by construction, and we note that we have rotated to the frame with $p_\perp$ = 0.  This also introduces a $\delta$-function setting $k^2=0$. 
After imposing all $\delta$-function constraints, one straightforward integral is left, for which we use dimensional regularization to remove its UV divergence, 
\begin{align}
D_{th}(z) = \alpha_y \int \dfrac{d^{2-2\epsilon} k_\perp}{(2\pi)^2} \dfrac{z\, k_\perp^2  + \frac{(1-z)^2}{z} M_t^2}{\left( k_\perp ^2  - \left( \frac{1-z}{z} \right) M_t^2 \right) ^2 + \left( \frac{1-z}{z} \right)^2 M_t^2 \, \Gamma_t^2}. 
\label{eq:dthintegrand}
\end{align}

We recognize that the $M_t^2$ term in the numerator, combined with the denominator, provides an ultracollinear correction to the LL result.  Additionally, the kinematics allow the top quark propagator to hit the pole at $(p+k)^2 = M_t^2$.  By working in the complex mass scheme\cite{Beneke:2003xh},\cite{Beneke:2004km}, and deforming $M_t \rightarrow M_t - i\, \Gamma_t$, we avoid a divergence, but do pick up a contribution enhanced by $1/\Gamma_t$.  At LO, $\Gamma_t = \frac{\alpha_y M_t}{4}$.  The physics of such a large correction makes sense since we are necessarily in the region of on-shell top production.  In addition to the usual perturbative splitting process that creates $b,\,h$ particles, the top can also decay to them, which it does with a probability that approaches 100\%.\footnote{If we were interested in the details of top decay product distributions near its mass pole, then it may be necessary to resum the large logs of $\Gamma_t/M_t$.  The Unstable Particle Effective Theory of \cite{Beneke:2015vfa} is one straightforward formalism for tackling this problem.  However, we take the observables from our processes of interest, jet production from energetic collider or astroparticle processes, as sufficiently inclusive that this resummation is beyond our scope.  It is also pointed out in Ref.~\cite{Manohar:2014vxa} that including finite widths breaks gauge invariance.  Gauge theories are beyond the scope of this work, but are an important future direction.}  
Upon performing the integral in Eq.~\ref{eq:dthintegrand} and renormalizing in $\overline{\text{MS}}$, we reach the final expression,
\begin{align}
D_{th}(z) = \dfrac{\alpha_y}{4\pi}\left[ z\, \log \left(\dfrac{\mu^2}{M_t^2} \right) - z\, \log\left( \frac{1-z}{z} \right) -1 + \dfrac{\pi\, M_t}{\Gamma_t} \right],
\end{align}
where $\mu$ is the renormalization scale.  The NLL terms, with no explicit $\log(\mu^2/M_t^2)$ or $1/\Gamma_t$ are due to the presence of finite $M_t$ in both numerator and denominator.  These are the ultracollinear contributions.  Lastly, the $1/\Gamma_t$ term modifies the running at LO from the $D^{-1}$ contribution to $\gamma_{ij}$ ({\it cf.}~Eq.~\ref{eq:gammadef}).  

It is also necessary to obtain the self-fragmentation functions and their running.  These are just the wavefunction renormalization factors for the fields in our theory.\footnote{Since the top is unstable, its pole and residue are complex valued.  Thus, the factor that enters the fragmentation function is $|\sqrt{Z_t}|^2$.  The imaginary part of $\sqrt{Z_t}$ only enters at $\mo(\alpha_y)$, and as we only work to that order, in practice we can drop it.}  Since these computations take a different form, we give an example here with $D_{hh}$ ($= Z_h\,\delta(1-z)$, where $Z_h$ is our $\overline{\text{MS}}$ wavefunction renormalization).  
\begin{align}
	 \begin{tikzpicture}[baseline=-0.35 ex]
\begin{feynman}
\vertex (a)[crossed dot] {};
\vertex [right=of a] (b);
\vertex [right=of b] (d);
\vertex [ right=of d] (e)[crossed dot]{};
\diagram* {
(a) -- [scalar,momentum=\(p\)](b),
(b)-- [fermion,half right, looseness=1,momentum'=\(p-k\)](d),
(b)-- [double,double distance=0.5ex,thick,with arrow=0.5,arrow size=0.15em,half left, looseness=1,momentum=\(k\)](d),
(d)-- [scalar,momentum=\(p\)](e)
};
\end{feynman}
\end{tikzpicture}
\end{align}
For the self-fragmentation function, the contribution of this graph to the residue to $\mathcal{O}(\alpha_y)$ is calculated as
\begin{align}
Z_h = \left( 1 - \dfrac{1}{p^+}\dfrac{d}{dp^-} \Sigma(p) \Big |_{p^2=0} \right)^{-1},
\end{align}
where $\Sigma(p)$ is the expression of the loop. We again compute in the SCET2 theory with dimensional regularization and $\overline{\text{MS}}$, obtaining
\begin{align}
D_{hh}(z)  = \left(1 - \dfrac{\alpha_y}{2\pi}\left[\log \left(\dfrac{\mu^2}{M_t^2} \right) +\frac 1 2 \right] \right) \delta(1-z)
\end{align}
Similar to the real emission graph, the $\log$ coefficient is the leading-log term.  The other $\mo(\alpha_y)$ contribution arises from the finite $M_t$ that we track in the SCET2 theory.  The calculations for the other real \& virtual-emission fragmentation functions proceed similarly.  In Appendix \ref{app:details}, one can find more calculational details for real and virtual fragmentation functions.  We collect the results for the entire theory in Table \ref{tbl:ffs} below.

In principle, there are other diagrams contributing to $D_{hh}$ and $D_{tt}$ of the following form\footnote{\relsize{0.7}{There are also two extra graphs, one contributing to $D_{hh}$ and one to $D_{bb}$, where they are the same but $t$ is substituted by $b$.}}:
\begin{align}
	D_{hh}: \begin{tikzpicture}[baseline=-0.35 ex]
\begin{feynman}
\vertex (a)[crossed dot] {};
\vertex [right=of a] (b);
\vertex [ right=of b] (e)[crossed dot]{};
\vertex [ above left=of b] (e1);
\vertex [ above right=of b] (e2);
\diagram* {
(a) -- [scalar](b),
(b)-- [double,double distance=0.5ex,thick,with arrow=0.5,arrow size=0.15em](e1),
(e2)-- [double,double distance=0.5ex,thick,with arrow=0.5,arrow size=0.15em] (b),
(e1)-- [double,double distance=0.5ex,thick,with arrow=0.5,arrow size=0.15em,half left, looseness=1](e2),
(b)-- [scalar](e)
};
\end{feynman}
\end{tikzpicture}
\qquad
D_{tt}: \begin{tikzpicture}[baseline=-0.35 ex]
\begin{feynman}
\vertex (a)[crossed dot] {};
\vertex [right=of a] (b);
\vertex [ right=of b] (e)[crossed dot]{};
\vertex [ above left=of b] (e1);
\vertex [ above right=of b] (e2);
\diagram* {
(a) -- [double,double distance=0.5ex,thick,with arrow=0.5,arrow size=0.15em](b),
(b)-- [scalar](e1),
(b)-- [scalar] (e2),
(e1)-- [scalar,half left, looseness=1](e2),
(b)-- [double,double distance=0.5ex,thick,with arrow=0.5,arrow size=0.15em](e)
};
\end{feynman}
\end{tikzpicture}
\end{align}
These graphs however, have trivial contribution at $\mo(\alpha_y)$ since $\dfrac{d}{dp^-}\Sigma(p)$ would give 0.

%%%%%%%%%%%%%%%%%%%%%%%%%%
\subsection{Summary of Results}
\label{subsec:ressum}
%%%%%%%%%%%%%%%%%%%%%%%%%%

We tabulate here all results that contribute to matching of Wilson coefficients at $\mo(\alpha_y)$ as we run from scales with $Q \gg M_t$ to energies below $M_t$, where the top is integrated out of the theory.  This includes shifts to the LO results from the finite top width, $\Gamma_t$.  We also include those parts of the anomalous dimensions that arise from the finite $M_t$ in the SCET2 regime and contribute to running at NLL order.  These are ultracollinear contributions in the sense that their existence requires the inclusion of nonzero $M_t$ in the initial phase of the calculation, but they persist if one takes $M_t \rightarrow 0$ at the end of it.  We dub these terms ``NLL-UC''.  We stress that we have not performed a full NLL resummation, as this requires two-loop results in the toy theory that are beyond our scope.  Nonetheless, in presenting numerical results in Section \ref{sec:res}, we show the shifts incurred in the Wilson coefficients by including the NLL-UC terms.  This is to demonstrate the size of the contributions that would be missed under a more naive treatment that keeps the top massless until it is integrated out.

\begin{table}[h!]
\centering
\begin{adjustbox}{width=0.5\textwidth}
\begin{tabular}[t]{|c|c|c|}
\hline
  &  &  \\
F.F. & Leading-Log & Finite $M_t$ corrections \\
 &  &  \\
\hline
 &  &  \\	
$D_{tb}\left(z\right)=$ & $\dfrac{\alpha_y}{4\pi} (1-z) \log\left(\dfrac{\mu^2}{M_t^2}\right)$ & $+\dfrac{\alpha_y}{4\pi}\left((1-z)\,\log\left(\dfrac{z}{1-z}\right)- 1 +\pi\dfrac{M_t}{\Gamma_t}\right)$\\
 &  &  \\
\hline
 &  &  \\
$D_{th}\left(z\right)=$ & $\dfrac{\alpha_y}{4\pi}\,z\, \log\left(\dfrac{\mu^2}{M_t^2}\right)$ & $+\dfrac{\alpha_y}{4\pi}\left(z\log\left(\dfrac{z}{1-z}\right) - 1 +\pi\dfrac{M_t}{\Gamma_t}\right)$\\
 &  &  \\
\hline
 &  &  \\
$D_{bt}\left(z\right)=$ & $\dfrac{\alpha_y}{2\pi} (1-z)\, \log\left(\dfrac{\mu^2}{M_t^2}\right)$ & $+\dfrac{\alpha_y}{2\pi}\left( (1-z) \, \log\left(\dfrac{z^2}{1-z}\right) + z \right)$\\
 &  &  \\
\hline
 &  &  \\
$D_{bh}\left(z\right)=$ & $\dfrac{\alpha_y}{2\pi}z \log\left(\dfrac{\mu^2}{M_t^2}\right)$ & $+\dfrac{\alpha_y}{2\pi}\left(z\log\left(z\right)  + 1 -z \right)$\\
 &  &  \\
\hline
 &  &  \\
$D_{hb}\left(z\right)=$ & $\dfrac{\alpha_y}{2\pi} \log\left(\dfrac{\mu^2}{M_t^2}\right)$ & $+\dfrac{\alpha_y}{2\pi}\left(\log\left(z\right)-1 +z\right)$\\
  &  &  \\
\hline
 &  &  \\
$D_{ht}\left(z\right)=$ & $\dfrac{\alpha_y}{2\pi}\log\left(\dfrac{\mu^2}{M_t^2}\right)$ & $+\dfrac{\alpha_y}{2\pi}\left(\log\left(\dfrac{z^2}{1-z}\right) -z \right)$\\
 &  &  \\
\hline
 &  &  \\
$D_{tt}\left(z\right)=$ & $\left[1-\dfrac{\alpha_y}{8\pi} \log\left(\dfrac{\mu^2}{M_t^2}\right)\right]\delta(1-z)$ & 0 \\
 &  &  \\
\hline
 &  &  \\
$D_{hh}\left(z\right)=$ & $\left[1-\dfrac{\alpha_y}{2\pi}\log\left(\dfrac{\mu^2}{M_t^2}\right)\right]\delta(1-z)$ & $-\dfrac{\alpha_y}{2\pi}\left(\dfrac{1}{2}\right)\delta(1-z)$\\
 &  &  \\
\hline
 &  &  \\
$D_{bb}\left(z\right)=$ & $\left[1-\dfrac{\alpha_y}{4\pi} \log\left(\dfrac{\mu^2}{M_t^2}\right)\right]\delta(1-z)$ & $-\dfrac{\alpha_y}{4\pi}\left(\dfrac{3}{2}\right)\delta(1-z)$\\
 &  &  \\
\hline
\end{tabular}
\end{adjustbox}
\caption{The fragmentation functions to  $\mathcal{O}(\alpha_y)$.  In the high-scale theory (SCET1), we only get the results in the LL column, and $M_t$ is strictly an IR cutoff on perp momentum.  In the intermediate-energy theory (SCET2), we pickup full $M_t$ dependence.  Since $\Gamma_t = \frac{\alpha_y M_t}{4}$, we see that all entries in the ``Finite $M_t$ corrections'' column survive in the $M_t \rightarrow 0$ limit, even though we need to keep $M_t \neq 0$ for their initial determination.}
\label{tbl:ffs}
\end{table}

As we will further see in Section \ref{sec:rm}, it is useful to write the Fragmentation Functions in matrix form:
\begin{align}
D_{ij}=\left(\begin{matrix}
	D_{hh} & D_{hb} & D_{ht}\\
	D_{bh} & D_{bb} & D_{bt}\\
	D_{th} & D_{tb} & D_{tt}
\end{matrix}\right)
\end{align}
In the high-scale SCET1 theory, where all particles are treated as massless, the LO ``fragmentation matrix'' is just $\delta(1-z)$ terms on the diagonal.  As discussed above, after passing to the intermediate-scale SCET2 theory with finite $M_t$ and $\Gamma_t$, the Fragmentation Functions at LO are modified to the following expression:
\begin{align}
D^{LO}_{ij}=\left(\begin{matrix}
	\delta(1-z) & 0 & 0\\
	0& \delta(1-z) & 0\\
	1 & 1 & \delta(1-z)
\end{matrix}\right)
\end{align}
The off-diagonal, $z$-independent terms arise from the near-certain decay of the unstable tops.  We give the full contributions up to NLO in $\mo(\alpha_y)$ in Table \ref{tbl:ffs}.
In Mellin Space, the total contribution in the intermediate-mass theory (SCET2) to NLO is
\begin{align}
D_{ij}&= D^{\text{LO}}_{ij} + D^{M_t=0,\,\text{NLO}}_{ij}\log\left(\dfrac{\mu^2}{M_t^2}\right) + D^{M_t\neq0,\,\text{NLO}}_{ij}=\nonumber\\
&=\left(
\begin{array}{ccc}
 1 & 0 & 0 \\
 0 & 1 & 0 \\
 \frac{1}{s} & \frac{1}{s} & 1 \\
\end{array}
\right)+\nonumber\\
&+\dfrac{\alpha_y}{2\pi}\left(
\begin{array}{ccc}
 -1 & \frac{1}{ s} & \frac{1}{s} \\[0.5em]
 \frac{1}{ (s+1)} & -\frac{1}{2} & \frac{1}{ s(s+1)} \\[0.5em]
 \frac{1}{2(s+1)} & \frac{1}{2  s (s+1)} & -\frac{1}{4 } \\
\end{array}
\right)\log\left(\dfrac{\mu^2}{M_t^2}\right) +\nonumber\\
&+\dfrac{\alpha_y}{2\pi}\left(
\begin{array}{ccc}
 -\frac{1}{2} & \frac{-2s-1}{ s^2 (s+1)} & \frac{H(s)}{s}-\frac{2}{s^2}-\frac{1}{s+1} \\[0.5em]
 \frac{1}{ s (s+1)^2} & -\frac{3}{4 } & \frac{s (s+1) H(s)+s^3-4s-2}{ s^2 (s+1)^2} \\[0.5em]
 \frac{H\left(s+1\right)}{2(s+1)}-\frac{1}{2(s+1)^2}-\frac{1}{2s} & \frac{s H(s)-(s+1)^2}{2  s^2 (s+1)} & 0 \\
\end{array}
\right),
\label{eq:dijmellin}
\end{align}
%\newpage
where $H(s)$ is the generalized harmonic number.  The Mellin-space variable $s$ is related to the momentum fraction, $z$, in the standard way, as shown in Eq.~\ref{eq:mellint}. For notational simplicity, we use the $D_{ij}$ notation for fragmentation functions in both Mellin ($s$) and momentum-fraction ($z$) space. One can easily obtain the high-scale (SCET1) fragmentation matrix from Eq.~\ref{eq:dijmellin} by dropping the off-diagonal elements in $D^{\text{LO}}_{ij}$, along with the $D^{M_t\neq0,\,\text{NLO}}_{ij}$ term.  As we discuss next, the running and log resummation necessary for showering particles in the different energy regimes of our theory is a `turn-the-crank' extension of the work described thus far.

%%%%%%%%%%%%%%%%%%%%%%%%%%%
\section{Running and Matching}
\label{sec:rm}
%%%%%%%%%%%%%%%%%%%%%%%%%%%

The explicit $\mu$-dependence we see in the renormalized fragmentation functions of Table \ref{tbl:ffs} lets us use RG evolution to run our Wilson coefficients down from their UV-determined boundary conditions to IR values.  In doing so, we automatically sum the LL contributions to our observable to all orders.  Indeed, one of the primary benefits of calculating in SCET is the straightforward way that this resummation follows from the RGE of its operators.  Recall from Figure \ref{fig:schem} that our calculation has four regimes: full Yukawa theory, SCET1 with $M_t=0$, SCET2 with finite $M_t$, and lastly a theory below $M_t$ with the top integrated out.  The full description thus involves three matching steps and two stages of RG running.  

The full-theory/SCET1 matching occurs at a high-scale we call $M_\chi$, thinking of it as the mass of a very particle that decays exclusively to tops.  The SCET1 running is standard and has no top mass effects by construction.  Then, at some scale $M_\text{intermediate}$, top particles are more likely than not to be produced on-shell and we must account for finite $M_t$.  Since $M_t$ is now comparable to our soft scale, we have a SCET2 theory.  The matching of Wilson coefficients from SCET1 is trivial.  However, for all scales below $M_\text{intermediate}$, they run with anomalous dimensions modified by finite $M_t$.\footnote{The SCET1/SCET2 transition is most simply handled by matching our SCET1 of interest to a fictitious SCET2 where the top mass is tuned by hand to be parametrically below its soft scale, and thus dropped.  Having obtained the Wilson coefficients of this artificial theory, we then tune $M_t$ to be larger until it coincides with the soft scale.  This necessarily modifies operator dimensions, but it cannot change the Wilson coefficients, which are set by UV physics and are thus decoupled from the soft scale, including $M_t$ effects.}  Lastly, in matching at $M_t$, one necessarily accounts for the finite-width and ultracollinear contributions due to $M_t$ in the fragmentation functions.

%%%%%%%%%%%%%%%%%%%%%%%%%%%%%%%%%%%%%
\subsection{RG Evolution for Wilson Coefficients and Fragmentation Functions}
\label{subsec:rg}
%%%%%%%%%%%%%%%%%%%%%%%%%%%%%%%%%%%%%

Although it is standard, we review here the evolution equation  derivation for fragmentation functions and their corresponding Wilson coefficients.  Our discussion largely follows that in \cite{Davidson:2020gsx}.  We will find that for an unstable particle, a careful accounting is needed even for a correct treatment at $\mo(1)$. If we take an observable that factorizes like our differential decay rate of interest ({\it cf.}~Eq.~\ref{eq:obsdef}), we can write it as follows, 
\begin{align}
	\sigma^\prime_j = C_i(\mu)D_{ij}(\mu),
\end{align}
where $\sigma^\prime_j$ is a differential rate involving particle $j$. Since this quantity must be $\mu$-independent,
\begin{align}
	\dfrac{d\sigma^\prime_j}{d\log(\mu)} = 0 \quad \Rightarrow \quad \dfrac{dC_i(\mu)}{d\log(\mu)}D_{ij}(\mu)+C_i(\mu)\dfrac{dD_{ij}(\mu)}{d\log(\mu)} =0
\label{eq:muind}
\end{align}
It is useful to first examine the fragmentation function; one can write the bare $D_{0,ij}$ in terms of the renormalized $D_{kj}$,
\begin{align}
	D_{0,ij} = Z_{ik}D_{kj},
\end{align}
where $Z_{ij}$ are the renormalization factors. This gives us an additional constraint to set up the RG equation since bare quantities are necessarily independent of the renormalization scale, 
\begin{align}
\dfrac{dD_{0,ij}}{d\log(\mu)} =0 \quad \Rightarrow \quad \dfrac{dZ_{ik}}{d\log(\mu)}D_{kj} + Z_{ik}\dfrac{dD_{kj}}{d\log(\mu)}=0.
\end{align}
Solving for $\dfrac{dD_{kj}}{d\log(\mu)}$ gives
\begin{align}
	\dfrac{dD_{mj}}{d\log(\mu)}=-Z^{-1}_{mi}\dfrac{dZ_{ik}}{d\log(\mu)}D_{kj}.
\end{align}
We thus recognize the anomalous dimension matrix,  $\gamma_{mk} \equiv Z^{-1}_{mi}\dfrac{dZ_{ik}}{d\log(\mu)}$, which gives
\begin{align}
	\dfrac{dD_{mj}}{d\log(\mu)} &= -\gamma_{mk}D_{kj} \Rightarrow \nn \\ 
 \dfrac{dD}{d\log(\mu)} &= -\gamma\cdot D
\end{align}
Since we can compute the renormalized fragmentation functions $D_{ij}$ straightforwardly, it is useful to invert the last equation to get
\begin{equation}
\boxed{\gamma=-\dfrac{dD}{d\log(\mu)}D^{-1} }
\label{eq:adeq}
\end{equation}
This form shows its utility in deriving the evolution equation for the Wilson Coefficients.  Rewriting the equation for $\mu$-independence of the observable ({\it cf.}~Eq.~\ref{eq:muind}), one gets
\begin{align}
\dfrac{dC_i}{d\log(\mu)}D_{ij}= - C_i\dfrac{dD_{ij}}{d\log(\mu)},
\end{align}
which allows us to introduce the anomalous dimension,
\begin{align}
\dfrac{dC_i}{d\log(\mu)}D_{ij}= - C_i\left(-\gamma_{ik}D_{kj}\right).
\end{align}
We can then eliminate $D_{ij}$ to get a flow equation for the Wilson coefficients phrased purely in terms of $C$ and $\gamma$, 
\begin{align}
\dfrac{d\,C_l}{d\log(\mu)} &= C_i\gamma_{il} \Rightarrow \nn \\  \Aboxed{\dfrac{d\,C}{d\log(\mu)} &= \gamma^T \cdot C}
\end{align}
As mentioned above, we perform our running in two distinct theories.  Even though they have the same particle content, their $\gamma$s differ at the $\mo(1)$ level due to the different scales they inhabit and whether or not one has a finite $M_t$.
In ``$z$-space'' where we work explicitly with plus-momentum fractions, we construct our observable as a convolution, 
\begin{align}
	\dfrac{1}{\Gamma_0}\dfrac{d\Gamma_{j+X}}{dz} = \int_z^1 \dfrac{dx}{x} C_i \left(x/z\right) D_{ij}\left(z\right).
\label{eq:obsdeftwo}
\end{align}
The running equations for Wilson Coefficients are then given by
\begin{align}
	\dfrac{dC_l(z)}{d\log(\mu)}=\int_z^1 \dfrac{dx}{x}  C_i(x/z)\,\gamma_{il} (z) 
\label{eq:crun} 
\end{align}
We can reduce this integro-differential equation to a mere differential equation by means of a  Mellin Transform, defined by the following relations,
\begin{align}
	&\left\lbrace\mathcal{M}f\right\rbrace(s) = \int_0^\infty z^{s-1} f(z) \qquad(\text{Mellin Transform})
\label{eq:mellint}
\\
&\left\lbrace\mathcal{M}^{-1}F\right\rbrace(z) = \dfrac{1}{2\pi i}\int_{c-i\infty}^{c+i\infty} z^{-s} F(s) \qquad(\text{Inverse Mellin Transform}),
\end{align}
with the property needed for simplifying convolution:
\begin{align}
	\left\lbrace\mathcal{M}\int\dfrac{dx}{x} \quad f(x/z) g(z)\right\rbrace(s) =  F(s) G(s)
\end{align}
Thus Eq.~\ref{eq:crun} transforms into a simpler algebraic equation,
\begin{align}
\dfrac{d\,C_l (s)}{d\log(\mu)} =  C_i(s)\gamma_{il}(s). 
\end{align}
Once again, for simplicity we use the same notation for functions and their (inverse) Mellin transforms.

%%%%%%%%%%%%%%%%%%%%%%%%%%%%%%%%%%%%%
\subsection{Running \& Matching in the Toy Yukawa Theory}
\label{subsec:rmyukawa}
%%%%%%%%%%%%%%%%%%%%%%%%%%%%%%%%%%%%%

We can now apply this general approach to evolving fragmentation functions and resumming large logs to our specific theory of interest.  We already have what we need in the Mellin space fragmentation function matrices given in Eq.~\ref{eq:dijmellin}.  To compute our low-scale observables, we now just need to apply the formalism of Section \ref{subsec:rg}.  We first compute the SCET2 LL anomalous dimensions, which are just the $\mo(\alpha_y)$ contributions. Using the notation of Eq.~\ref{eq:dijmellin}, we have
\begin{align}
	\gamma^{\text{LL}}_{ij}&= -\dfrac{d D_{ik}}{d \log(\mu)} D_{kj}^{-1} \Big |_{\mo(\alpha_y)}= - 2D_{ik}^{m=0,\,\text{NLO}}	\left(D_{kj}^{\text{LO}}\right)^{-1}=\nonumber\\
	&= \dfrac{\alpha_y}{\pi}\left(\begin{matrix}
 -\delta(1-z)& 1 &  1\\
z & -\dfrac{1}{2}\delta(1-z) & 1-z \\
\dfrac{z}{2}  & \dfrac{(1-z)}{2} & -\dfrac{1}{4}\delta(1-z)
\end{matrix} \right)\left(
\begin{array}{ccc}
 \delta(1-z) & 0 & 0 \\
 0 & \delta(1-z)& 0 \\
 1 & 1 & \delta(1-z) \\
\end{array}
\right)^{-1}.
\label{eq:llad}
\end{align}
A simple way to handle the inverse matrix is to convert everything to Mellin space, and then transform back to $z$ space.  It is straightforward to get the LL SCET1 anomalous dimensions from Eq.~\ref{eq:llad}.  We just need to zero out the off-diagonal entries in $D_{kj}^{\text{LO}\; -1}$. These terms arise from the $\Gamma_t$ that follows from having finite $M_t$.  The operator running thus undergoes an $\mo(1)$ shift once our scale is sufficiently low that we evolve nearly on-shell particles. 

Running at NLL just requires us to calculate $\gamma_{ij}$ from Eq.~\ref{eq:adeq} to $\mo(\alpha_y^2)$.  However, this  result requires a full accounting of two-loop diagrams (equivalently, three-body phase space for real emissions) and is beyond our scope.  The ultracollinear terms in $D_{ij}$, those that are $M_t$ independent, but whose existence depends on keeping $M_t$ in the initial calculation, do contribute to $\gamma_{ij}$ at $\mo(\alpha_y^2)$.  When we present results in Section \ref{sec:res} we show Wilson coefficients that have undergone ``NLL-UC'' running, which includes just the ultracollinear contributions at subleading order.  This is necessarily incomplete, and the results presented below are not meant to represent a systematic approximation to the true NLL result.  Nevertheless, we see that the size of the ultracollinear contributions at NLL is non-negligible, and thus an important numerical effect {\it would be missed} if one were to drop fermion masses in a regime where one should account for them.\footnote{Naively computing $\gamma_{ij}$ to $\mo(\alpha_y^2)$ from our one-loop results in Table \ref{tbl:ffs} generates some terms of the form $\alpha_y^2 \log(\mu)$.  In running the Wilson coefficients, these would integrate to $\alpha_y^2 \log^2$ contributions.  However, that is an order that one gets from exponentiating the LL result.  If it were to survive, it would imply that the LL resummation fails.  Since that is not the case, as it would violate the $\mu$-independence of our observable, a complete calculation of $\gamma_{ij}$ to $\mo(\alpha_y^2)$ must cancel any contributions of the form $\alpha_y^2 \log$.  Thus, we drop any such terms in our NLL-UC anomalous dimensions.}  

The $\mo(\alpha_y^2)$ contribution to the anomalous dimension matrix that captures the ultracollinear terms is:
\begin{align}
	\gamma^{\text{NLL-UC}}_{ij} &= -\dfrac{d D_{ik}}{d \log(\mu)} D_{kj}^{-1} \Big |_{\mathcal{O}(\alpha_y^2),(\alpha_y^2 \log(\mu)) \tor 0} \nn \\
	&=- \left(
 \dfrac{d\alpha_y}{d\log\mu}\dfrac{\partial D_{ik}^{m\neq0,\text{NLO}}}{\partial \alpha_y}+2D_{ik}^{m=0,\text{NLO}}\right)	\left(D_{ik}^{\text{LO}} + D_{ik}^{m\neq0,\text{NLO}}\right)^{-1} \Bigg |_{\mathcal{O}(\alpha_y^2)}  \nn \\
	&=-\left(\dfrac{d\alpha_y}{d\log\mu}\dfrac{\partial D_{ik}^{m\neq0,\text{NLO}}}{\partial \alpha_y}\cdot D_{kj}^{-1\,\text{LO}}-2D_{ik}^{m=0,\text{NLO}} \cdot D_{kj}^{m\neq 0,\text{NLO}} \right), 
\label{eq:nllucad}
\end{align}
where we again classify terms using the terminology of Eq.~\ref{eq:dijmellin}.  We recognize in Eq.~\ref{eq:nllucad} factors of the $\beta$-function of our coupling, which is just
\beq
\beta(\alpha_y) \,=\, \frac{7 \alpha_y^2}{4\pi}.
\eeq
The Mellin transform of the matrix given by Eq.~\ref{eq:nllucad} is a complicated expression that we give in Appendix \ref{app:ad}. Nonetheless, it is straightforward to construct.

Having obtained what we need to calculate the effects due to running, we can proceed to consider matching.
The basic idea is that  our observable, $\dfrac{1}{\Gamma_0}\dfrac{d\Gamma_j}{dz}$ is the same in both theories at the matching scale.
\begin{align}
	\dfrac{1}{\Gamma_0}\dfrac{d\Gamma_j(\mu=\mu_M+\epsilon)}{dz}=\dfrac{1}{\Gamma_0}\dfrac{d\Gamma_j^\prime(\mu=\mu_M-\epsilon)}{dz}
\end{align}
where $j$ labels the observed particle, the LHS represents a higher-scale theory and the RHS is the lower-scale one. 
Using the factorized form of our differential decay rate ({\it cf.}~\ref{eq:obsdef}) and going to Mellin space, we have
\begin{align}
	 C_i(s,\mu_M) D_{ij}(s,\mu_M) = C'_k(s,\mu_M) D'_{kj}(s,\mu_M) 
\label{eq:matchop}  
\end{align}
We note that this equation holds $j$ by $j$, and thus that particle must necessarily exist in both theories. 
Solving for $C_k'$, we get 
\begin{align}
	C'_k(s,\mu_M)= 
    C_i(s,\mu_M) D_{ij}(s,\mu_M) \left(D'_{jk}(s,\mu_M) \right)^{-1},
\end{align}
matching order by order in $\alpha_y$. 

In the tower of theories that takes us from the creation of our massive fermion, the top, to scales below its mass, where we integrate it out, we pass through three matching scales ({\it cf.}~Fig.~\ref{fig:schem}).  We provide here an overview of the issues with each of these.  In Section \ref{sec:res}, we present a specific numerical implementation.

\begin{itemize}
\item \underline{Matching between the Full Theory and SCET1 at $M_\chi$:}\\

The main nontrivialites at this stage are the complications that arise when passing from a conventional quantum field theory to SCET.  However, this is now a standard procedure involving a well-established effective field theory \cite{Bauer:2000ew,Bauer:2000yr,Bauer:2001ct,Bauer:2002uv}.  One simplification relative to the original QCD SCET is that we have no issues with gauge invariance in a pure Yukawa theory.  By construction, the kinematically-enhanced fragmentation region of a collinear splitting of the top falls out of the leading-power SCET Feynman rule.  It is this matching into SCET that sets the UV boundary conditions for our Wilson coefficients that will undergo further running.  At LO, only tops arise from the $\chi$ decay.  However, at NLO it becomes possible to produce bottom or higgs as a hard, three-body decay, a production mechanism not captured the fragmentation of a top with the SCET vertex.  Thus, matching to our observable, $(1/\Gamma_0)\, d\Gamma/dz$, we find
\begin{align}
C_t(z,\mu=M_\chi) &= \delta(1-z) \nn \\
C_b(z,\mu=M_\chi) &= \alpha_y \left(\frac{3 z-2}{8 \pi } \right) \nn \\
C_h(z,\mu=M_\chi) &= \alpha_y\frac{z \left(-3 z^2+z+2\right)+2 \log (1-z)}{8 \pi  z^2}.
\label{eq:sopmatch}
\end{align}

\item \underline{Matching between SCET1 and SCET2 at $M_\text{Intermediate}$:}\\

In principle, there are many shifts that can occur in passing from a SCET1 to a SCET2 theory.  The former has a hierarchy between the virtualities of its soft and collinear sectors, and the latter does not.  Additionally, the change in power counting of the soft sector also shuffles the (ir)relevance of interactions.  Furthermore, the SCET2 theory is sensitive to the finite value of $M_t$ by construction.  Nevertheless, the $C_i$ Wilson coefficients of our fragmentation operators enjoy many simplifications.  As discussed above, the UV physics that sets $C_i$ is decoupled from $M_t$.  In practice, we match to a fictitious SCET2 with $M_t=0$, and then tune $M_t$ up to the SCET2 soft scale, which can have no effect on the $C_i$ we obtained.  Additionally, our fragmentation functions are computed from collinear fields alone.  The SCET1 soft contributions are removed by zero-bin subtraction \cite{Manohar:2006nz}, while the SCET2 soft vertex is power-suppressed.  This means that we are matching $D_{ij}$ computed in the collinear regime of both theories with $M_t=0$.  Thus, we have 
\begin{align}
	 C_i(s,M_\text{Intermediate}) D_{ij}(s,M_\text{Intermediate}) = C'_k(s,M_\text{Intermediate}) D'_{kj}(s,M_\text{Intermediate}), 
\end{align}
with the same $D$-matrices on both sides.  Thus, we get $C_i = C_i^\prime$, {\it i.e.}~a trivial matching.  It is the case though, that the NLL running in the SCET2 theory will be immediately different from that in SCET1 due to the finite $M_t$ effects.  

\vspace{0.1in}

\item \underline{Matching between SCET2 and the Low-Scale Theory at $M_t$:}\\

In some sense, this last matching is quite conventional; we integrate out a particle below its mass.  In the context of fragmentation, our matching must account for the fact that the $t$ had been a propagating source of $h,b$ particles, and now is gone from the theory.  This is where the finite-width contributions to $D_{th}$ and $D_{tb}$ come in ({\it cf.}~Table \ref{tbl:ffs}).  In matching to the Low Scale Theory with no $t$, they give one last, $\mo(1)$ injection of $h,b$ as we account for every $t$ particle's decay into them.  Once in the Low-Scale Theory (LST), interactions only proceed through irrelevant operators.  We thus get no leading-power contributions to fragmentation.  Thus, we have a simple fragmentation matrix,  
\begin{align}
D_{ij\,\text{LST}}(s,\mu=M_t) =\left(\begin{matrix}
1  & 0 \\
0 & 1 
\end{matrix} \right).
\end{align}

Since we only need to match $h,b$ final states, we only need the SCET2 fragmentation matrix with the $h,b$ columns.  To NLO in Mellin space, these are:
\begin{align}
D_{ij,2}(z)(s,\mu=M_t) =\left(\begin{matrix}
1  & 0 \\
0 & 1 \\
\frac{1}{s} & \frac{1}{s}
\end{matrix} \right) + \dfrac{\alpha_y}{8\pi} \left(
\begin{matrix}
 -2 & -4\frac{2 s+1}{s^3+s^2} \\
 \frac{4}{s (s+1)^2} & -3 \\
 2\left(\frac{H\left(s+1\right)}{s+1}-\frac{1}{s}-\frac{1}{(s+1)^2} \right)& 2\left(\frac{\frac{s H(s)}{s+1}-s-1}{s^2} \right)
\end{matrix} \right), \nn \\
\end{align}
The triviality of $D_{ij\,\text{LST}}$ lets us easily compute the $C_{i\,\text{LST}}$ from Eq.~\ref{eq:matchop}.

\end{itemize}

The scale $M_t$ is the lowest in our toy theory, and in a sense we are done.  The $b$ and $h$ fields in the perturbative Low-Scale Theory are interpolating operators for its $b$ and $h$ asymptotic states.  Much of our physical case of interest in the indirect detection of dark matter involves perturbative particles like photons and electrons.  In this case, the identification of interpolating operators that overlap with the physical asymptotic states is similarly straightforward.  One may wonder though, about the production of hadronic states.  For these cases, our fragmentation functions will eventually need to interface with a shower Monte Carlo that includes hadronization like Pythia \cite{Sjostrand:2014zea} or Herwig \cite{Bahr:2008pv}, as is done with PPPC \cite{Marco_Cirelli_2011} or HDMSpectra.\cite{Bauer:2020jay}

%%%%%%%%%%%%%%%%%%%%%%%%%%%
\section{Numerical Results \& Conclusions}
\label{sec:res}
%%%%%%%%%%%%%%%%%%%%%%%%%%%

We have seen the subtle and important physics of tracking a fermion's mass as one propagates it down in scale from its production to its eventual decay.  It is nevertheless useful to provide a numerical example to see how these effects shift results quantitatively.  In keeping with our Standard Model inspiration, we take $M_t$ = 173 GeV and we set $y_t(M_t)= 1$.  We make the $b$ and $h$ massless.  For describing fragmentation, at scales immediately above $M_t$, we work in SCET2.  

We can ask though, at what scale we can stop tracking $M_t$ and instead pass to the SCET1 theory with $M_t=0$.  As follows from our discussion in Section \ref{sec:intro}, our condition to account for the fermion mass is the scale below which its ``no-split'' probability, {\it i.e.}~the odds that we have produced a physical, on-shell particle are greater than 50\%.  We compute the Sudakov (no-split) factor for our $t$ following the presentation in \cite{Schwartz:2014sze}.  That reference defines the splitting function, $P_{ij}$, as the $z$-space anomalous dimension with $\alpha_y/\pi$ stripped off.  Inverse Mellin transforming $\gamma^\text{LL}_{th}$ from Table \ref{tab:gammall}, we get $P_{th} = 1/4 + z/2$. Thus, the Sudakov factor is 
\beq
\Delta(Q,M_t) = \exp \left( -\int_{M_t^2}^{Q^2} \frac{\alpha_y(\sqrt{t})}{4\pi} \frac{1}{t} dt  \right).
\eeq
For the above values of $y_t,M_t$, we find that the value of $Q$ for which $\Delta(Q,M_t) = 0.5$ is $Q = 1.5 \times 10^8$ TeV.\footnote{As an alternative to the use of Sudakov factors, one can do a consistency check within the fragmentation formalism itself.  We can make use of the probabilistic form of fragmentation functions at LL ({\it cf.}~\ref{eq:ffconserv}), and take the coefficient of $D_{tt}$ as the probability for the top not to split.  If we track where this is $\approx$0.5 after running to a high scale, we also get $\sim10^8$ TeV, in agreement with the simpler Sudakov determination.  In this case, $M_t$ provides a manifest physical IR regulator.  There are no fragmentation functions involving $t$ at scales below $M_t$, as we work in the Low-Scale Theory there, with the top integrated out.}  We thus take this scale as the upper limit of our SCET2 theory.  This may seem like a shockingly high scale at which to still be tracking finite $M_t$.  After all, we are nine orders of magnitude larger and $(M_t/Q)^2$ effects are beyond negligible.  However, the lesson of the ultracollinear terms is that naive power suppression breaks down as kinematic enhancements can provide compensating factors.  Said differently, in the limit that $\alpha_y \tor 0$, even if $M_\chi \gg M_t$, the $t\bar t$ pair the $\chi$ decays into will necessarily be on-shell as they will have no interactions that allow them to split and shower into other particles.  Furthermore, since our Yukawa theory has only collinear, rather than soft and collinear divergences like gauge theories, the Sudakov factor is only single-log enhanced.  This causes a relative suppression in the exponential argument, causing an extension of the regime where we need to track $M_t$.  

For the case of real QCD, if we use the double-log $q \tor gq$ Sudakov in \cite{Schwartz:2014sze} and plug in the observed values for top and bottom quark masses ($m_t,m_b$) along with $\alpha_s$, we find that one should keep bottom quark masses in calculations up to $Q \approx$ 235 GeV and top quark masses up to $Q \approx$ 43 TeV.  Returning to our toy theory, we summarize the parameters used in our example in Table \ref{tbl:params}.  The interesting mass effects occur in the SCET2 theory and its matching to the Low-Scale Theory.  Thus, to simplify our example we take the SCET1 regime to run over a negligible range of scales.  In effect, we match directly from the full theory to the SCET2 theory, since the Wilson coefficients we obtain are independent of $M_t$.\footnote{The boundary conditions we set in Eq.~\ref{eq:sopmatchmellin} are thus taken as the initial boundary conditions for the SCET2 running, {\it i.e.}~$C_i(M_\chi) = C_i(M_\text{Intermediate})$.}  We do not need to specify a numerical value for $M_\chi$, but we take it to be just a small factor greater than the $M_\text{Intermediate} = 1.5 \times 10^8$ TeV, determined by our Sudakov condition.  

As discussed in Section \ref{subsec:rmyukawa}, matching to the full theory at NLO gives $\mo(\alpha_y)$ contributions to $C_{h,b}$ since it is possible to produce $h,b$ particles from a hard three-body decay of the $\chi$.  By construction, these are the portions of the $\chi$ decay rate that do not follow from producing a $t$, which subsequently fragments into the other fields following the SCET Feynman rules.  We give the $z$-space boundary conditions in Eq.~\ref{eq:sopmatch}, but in practice we run and match in Mellin space.  Those boundary conditions are:
\begin{align}
C_t(s,\mu=M_\chi) &= 1 \nn \\
C_b(s,\mu=M_\chi) &= \alpha_y \left(\frac{s-2}{8 \pi  s (s+1)}\right) \nn \\
C_h(s,\mu=M_\chi) &= \alpha_y \left(-\frac{H(s-2)}{4 \pi  (s-2)}+\frac{1-5 s}{8 \pi  \left(s-s^3\right)}\right).
\label{eq:sopmatchmellin}
\end{align}
Presenting results is more physically transparent back in $z$ or plus-momentum fraction space.  Thus, after numerically solving our RG equation to run and/or match our Wilson coefficients at the scale $\mu=M_t$, we perform an inverse Mellin transform.  We set our contour as $s(t) = 1 + t\cdot\exp(i\,\phi)$, where $\phi = 3\pi/4$, and integrate numerically over $t$,
\begin{align}
C_i(z) = \text{Im}\left\lbrace\dfrac{1}{\pi}\int_{0}^{\infty}e^{i\phi} z^{-s(t)} C_i\left(s(t)\right) dt \right\rbrace,
\end{align}
We tested robustness of the results in shifting the definition of $s(t) = a + t\cdot\exp(i\,\phi)$, varying $a$ and $\phi$ by $\mo(1)$ amounts.  Solving the RG equation necessitates running the coupling, $\alpha_y$, as well,
\begin{align}
	\beta(\alpha_y) = \dfrac{7}{4\pi}\alpha_y^2; \qquad \alpha_y (\mu) = \dfrac{4\pi}{16\pi^2 - 7\log(\dfrac{\mu}{M_t})}
\end{align}
where we use $\alpha_y(M_t) = 1/4\pi$ as a boundary condition for $\alpha_y(\mu)$.  

\begin{table}[h!]
\centering
\begin{adjustbox}{width=0.9\textwidth}
\begin{tabular}[t]{|c|c|c|}
\hline
  &  &  \\
Energy Scale & EFT & Boundary Conditions \\
 &  &  \\
\hline
&  &  \\	
$\gtrsim$ Few $\times 10^8$ TeV & Full Theory &  \\
 &  &  \\
\hline
 &  &  \\	
Few-1.5 $\times 10^8$ TeV & SCET1 & At $M_\chi = $ Few $\times 10^8$ TeV; $C_t = \delta(1-z)$; $C_{b,h}=${\it cf.}~Eq.~\ref{eq:sopmatch} \\
 &  &  \\
\hline
 &  &  \\
1.5 $\times 10^8$ TeV-$M_t(=173\;\text{GeV})$  & SCET2 & At $M_\text{Intermediate} = 1.5 \times 10^8$ TeV; Trivial from matching SCET1 onto SCET2 \\
 &  &  \\
\hline
 &  &  \\
Below $M_t$ & Low-Scale Theory (No running) & At $M_t$; Matching Conditions that remove $t$ \\
 &  &  \\
\hline
\end{tabular}
\end{adjustbox}
\caption{Numerical inputs \& operational steps for computing with our toy theory.}
\label{tbl:params}
\end{table}

%%%%%%%%%%%%%%%%%%%%%%%%%%%%%%%%%%%%%%%%%%%%%%%%%
\subsection{Effect of Mass on Running}
\label{subsec:massrun}
%%%%%%%%%%%%%%%%%%%%%%%%%%%%%%%%%%%%%%%%%%%%%%%%%

\begin{figure}[H]\centering
\begin{minipage}{\textwidth}
\centering
\includegraphics[width=0.4\linewidth]{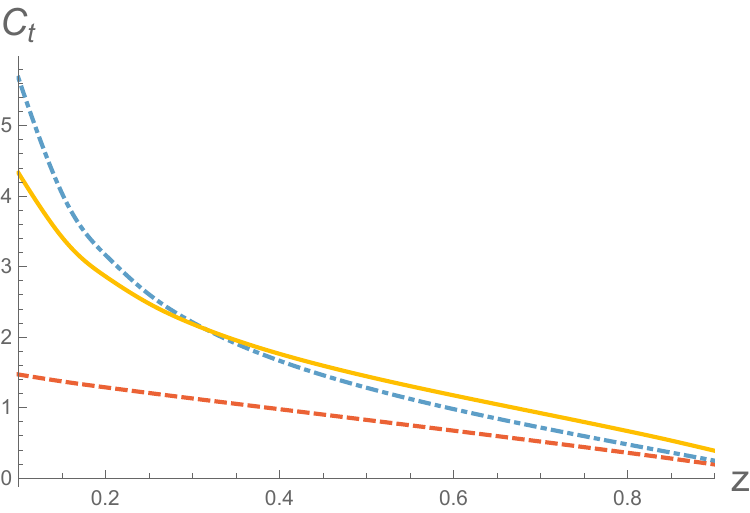}
\end{minipage}\hfill \\
\hfill\\ \hfill\\
\hfill\\
\hfill
\begin{minipage}{.45\textwidth}
\centering
\includegraphics[width=\linewidth]{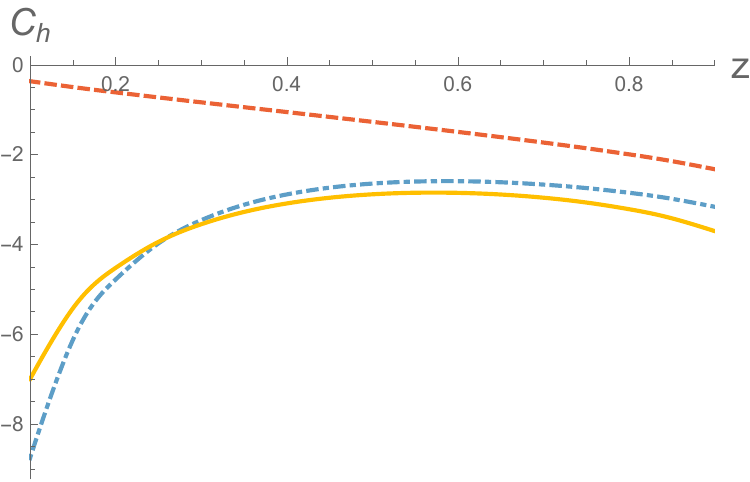}
\end{minipage}\hfill
\begin{minipage}{.45\textwidth}
\centering
\includegraphics[width=\linewidth]{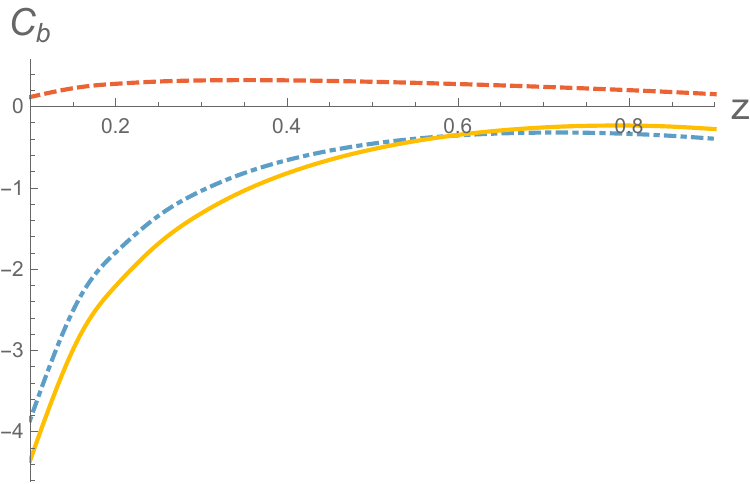}
\end{minipage}
\caption{Running of $C_t (z), C_h (z), C_b (z)$ at massless LL \textit{(dashed red)}, at LL including the finite width \textit{(dotted dashed blue)}, and at NLL-UC \textit{(dotted grey)}.  The NLL-UC results just add the $\mo(\alpha_y^2)$ contributions arising from ultracollinear terms to the anomalous dimensions.  We plot them just to demonstrate the size of the shifts they contribute to the $C_i$. The resulting NLL-UC curves are not meant to give a close approximation of the full NLL result.}
\label{fig:run}
\end{figure}

In Figure \ref{fig:run} we show the numerical significance of the finite $M_t$ contributions by comparing three different versions of running the Wilson Coefficients. Firstly, we have the massless case where no mass contributions are considered throughout our running.  That is to say that we include neither finite-width ($\Gamma_t$) nor ultracollinear contributions.  This is just LL conventional running with the usual splitting functions as obtainable from \cite{Ciafaloni:2010ti}. Said differently, this is what the running would be if we had a massless SCET1 theory from $M_\text{Intermediate}$ to $M_t$.  In the second case, we include the finite width effect, which enters the LL anomalous dimension ({\it cf.}~Eq.~\ref{eq:llad} and Table \ref{tab:gammall}). This is the case spelled out in Table \ref{tbl:params} with a SCET2 theory involving a finite $M_t$ running over our scales of interest.  We see that the effect is numerically large, as we expect since the finite width shifts the $\mo(\alpha_y)$ anomalous dimension at this same level.  This does not involve however, any of the ultracollinear contributions.  To see their effect we plot a third curve, which includes a subset of the $\mo(\alpha_y^2)$ anomalous dimension corrections ({\it cf.}~Tables \ref{tab:gammanllucb}-\ref{tab:gammanlluct}).  These give our ``NLL-UC'' running, which is not a complete accounting of NLL, as discussed above.  Nonetheless, we see the ultracollinear terms on their own give a non-negligible shift to the $C_i$.  Computing the other NLL contributions is beyond our scope.  Even if they largely cancel the shifts due to our NLL-UC terms plotted in Figs.~\ref{fig:run} and \ref{fig:match}, we nonetheless have an interesting result.  {\it Failing} to include NLL-UC by neglecting the finite fermion mass would spoil this cancellation, and one would thus obtain an erroneously large shift at NLL.  Without doing the full calculation, one cannot say for sure whether NLL-UC is important because it contributes to a non-negligible numerical shift from LL or because it participates in a nontrivial cancellation that leads to a {\it small} shift with respect to LL, but either way it plays an essential role in correcting the answer.
% 

%%%%%%%%%%%%%%%%%%%%%%%%%%%%%%%%%%%%%%%%%%%%%%%%%
\subsection{Effect of Mass on Matching}
\label{subsec:massmatch}
%%%%%%%%%%%%%%%%%%%%%%%%%%%%%%%%%%%%%%%%%%%%%%%%%

\begin{figure}[H]
\centering
\begin{minipage}{.45\textwidth}
\centering
\includegraphics[width=\linewidth]{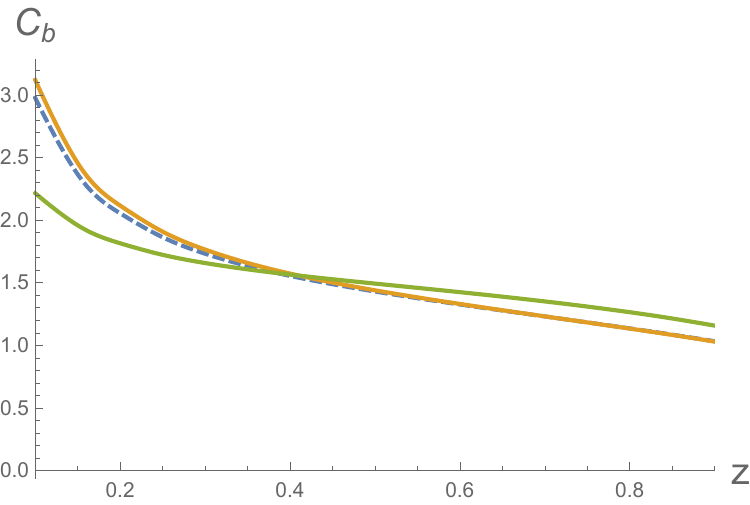}
\end{minipage}\hfill
\begin{minipage}{.45\textwidth}
\centering
\includegraphics[width=\linewidth]{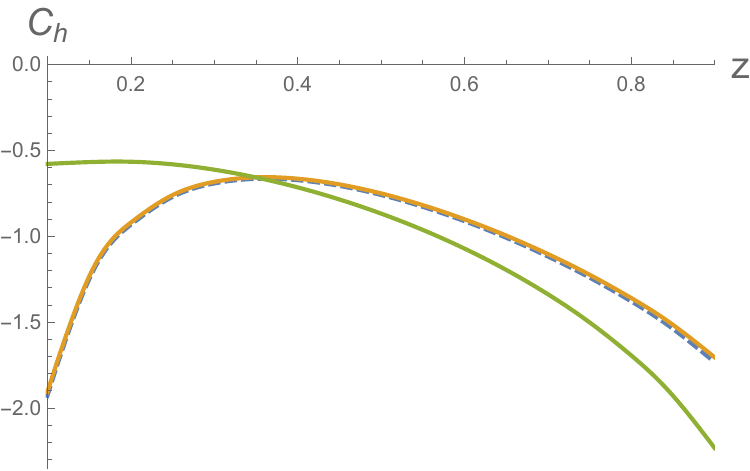}
\end{minipage}
\caption{Results of $C_h (z), C_b(z)$ in Running + Matching at LL+LO \textit{(dashed blue)}, LL+NLO \textit{(solid orange)}, NLL-UC+NLO \textit{(dotted grey)}.  The NLL-UC results just add the $\mo(\alpha_y^2)$ contributions arising from ultracollinear terms to the anomalous dimensions.}
\label{fig:match}
\end{figure}

\begin{figure}[H]
\centering
\begin{minipage}{.45\textwidth}
\centering
\includegraphics[width=\linewidth]{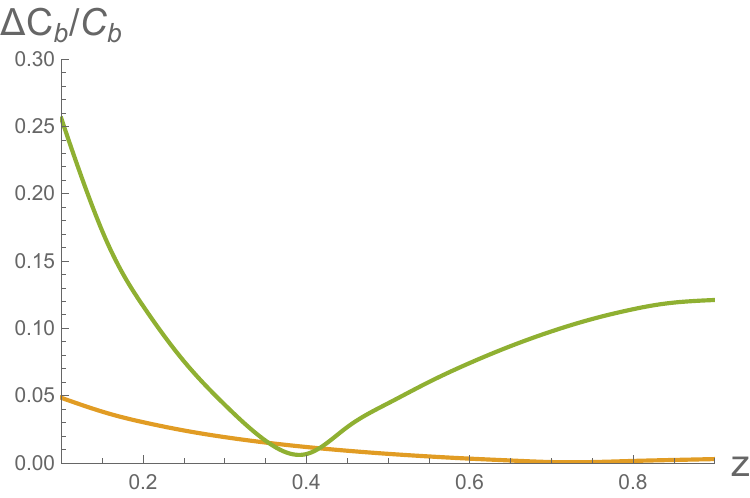}
\end{minipage}\hfill
\begin{minipage}{.45\textwidth}
\centering
\includegraphics[width=\linewidth]{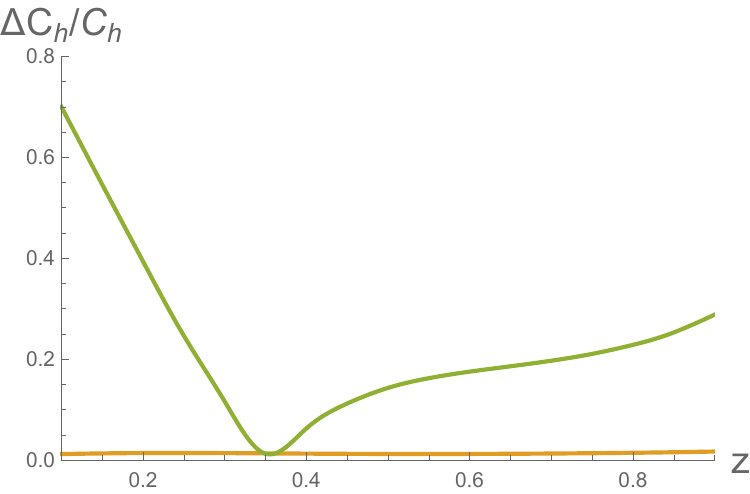}
\end{minipage}
\caption{ Ratio of $\dfrac{|C_\text{NLO}^\text{LL} - C_\text{LO}^\text{LL}|}{C_\text{LO}^\text{LL}}$ \textit{(orange)} and $\dfrac{|C_\text{NLO}^\text{NLL-UC} - C_\text{LO}^\text{LL}|}{C_\text{LO}^\text{LL}}$ \textit{(dotted grey)} for $C_b, C_h$.}
\label{fig:rats}
\end{figure}

If we were generating particles in this $t,b,h$ theory originating from some high-energy process (which we take to be the decay of some heavy $\chi$ for concreteness), running down in scale would eventually drop below $M_t$.  This is the final step of Table $\ref{tbl:params}$, matching to a theory without the $t$.  In Fig.~\ref{fig:match}, we plot the result for the two remaining Wilson coefficients, $C_b, C_h$. 
Once again, we show three variations.  In all cases, we assume that the UV theory in this matching is SCET2 with finite $M_t$.  Thus, the LL running does account for finite $\Gamma_t$.  By matching at LO, we again include the finite width, which is unavoidable even in a leading treatment with a massive particle.  At NLO, we include the ultracollinear terms in the fragmentation functions ({\it cf.}~Table \ref{tbl:ffs}).  We see in Fig.~\ref{fig:rats}, this just provides a percent-level shift.  However, in both Figs.~\ref{fig:match} and \ref{fig:rats}, we see that running at NLL-UC provides a qualitatively larger shift.  As discussed above, this shift may or may not persist upon doing full NLL resummation, but a sizeable error would be made in $C_{b,h}$ if one neglected the ultracollinear terms entirely.  It is an interesting question for future work if their effects survive in the complete NLL result or get substantially cancelled by the other $\mo(\alpha_y^2)$ contributions to $\gamma_{ij}$.

%%%%%%%%%%%%%%%%%%%%%%%%%%%
\subsection{Conclusions}
\label{subsec:conc}
%%%%%%%%%%%%%%%%%%%%%%%%%%%

From a rather ordinary mathematical problem arises a very interesting physical paradox that requires nontrivial conceptual labor to resolve. As Smilga noticed long ago \cite{Smilga:1990uq}, small fermion mass terms are not obeying commutation of the massless limit, and thus it is unsafe to simply toss them aside. The use of Effective Field Theory and its systematic formalism of operators and Wilson coefficients reveals when one must track a fermion mass.  The answer boils down to when one expects to produce an on-shell particle, as quantified by the Sudakov factor.  It is not hard to cook up an example, as we have done here, where finite mass effects endure long beyond the naive intuition of $(m_\text{light}/m_\text{heavy})^2$ suppression.

We see furthermore that the numerical stakes of keeping finite particle masses (or not) can be significant.  There are $\mo(\text{few})$ shifts in the Wilson coefficients, $C_i$, depending on whether one tracks the finite width in running, with further possible $\mo(10\text{s}\%)$ corrections from the ultracollinear contributions to the anomalous dimensions ({\it cf.}~Fig.~\ref{fig:run}).  Similar scale shifts from these terms persist in our final matching step, where we integrate out the massive fermion ({\it cf.}~Figs.~\ref{fig:match} \& \ref{fig:rats}). 

Now that we have studied and understood the subtleties of finite fermion mass effects in a controlled setting, it is a natural follow-up to study more realistic theories.  Our quick exercise with QCD Sudakovs in Section \ref{sec:res} shows that even with double-log running, one may need to track $b$ and $t$ quark masses about an order of magnitude higher than one might expect.  For showers used for astroparticle physics, like HDMSpectra \cite{Bauer:2020jay}, the range of scales under consideration is enormous, stretching from weak to Planck scale.  One could imagine all manner of BSM physics in this range.  For each particle, we must answer the question of when to keep its mass.  By extending this work, one can arrive at a systematic answer.

\vspace{0.3in}

{\it Acknowledgments:}
This work has benefitted from useful discussions with Benoit Assi, Sean Fleming, Stefan Hoeche, Aneesh Manohar, Thomas Mehen, Davison Soper, and Ubirajara van Kolck.  We especially thank Varun Vaidya for clarifying subtle points in SCET.  We were grateful to have Ozan Erdo\u{g}an's collaboration during the early stages of this project.
MB and PC are supported by the DOE (HEP) Award DE-SC0019470.

\appendix

%%%%%%%%%%%%%%%%%%%%%%%%%%%
\section{Fragmentation Formalism}
\label{app:ff}
%%%%%%%%%%%%%%%%%%%%%%%%%%%

We review here the fragmentation formalism and how our calculational quantities of interest arise.  From SCET, we take the operator definitions in \cite{Procura:2009vm,Jain:2011xz}, which follow from the classic treatment in perturbative QCD given in \cite{Collins:1981uw}.  As originally formulated, the physical concern was observing hadrons in jets initiated by perturbative partons.  Our situation here is qualitatively different, as the theory is fully perturbative.  Nonetheless, we can still work with fragmentation functions, as in \cite{Bauer:2020jay}, which provide a means to compute the rate to observe a particle $j$ that radiated from a possibly different particle $i$.    

It is natural to conceive of the fragmentation function as a kind of probability distribution.  In its original QCD framing, this would give the probability to observe a hadron with a particular energy fraction of its mother parton.  In our perturbative theory, this becomes the probability of observing a ``parton'' with a particular plus-momentum fraction of its mother parton.  The probabilistic interpretation breaks down beyond the Leading-Log level.  One can get explicitly negative results, in fact.  While inconvenient, this not a fundamental pathology.  Only observable amplitudes need to respect the strict unitarity of quantum mechanics.  Nonetheless, having even an approximately probabilistic quantity is useful for guiding its construction.  Also, the need to conserve probability at LL provides a useful sum rule with which to check the calculation, as detailed below. 

Following Collins, as reviewed in \cite{Collins_2011}, constructing a function which gives the number density of particles with a particular fraction of their mother's $E$ or $p^+$ starts us in a frame where the mother has zero transverse momentum, ($k_\perp$ = 0), and we integrate over the daughter's $\vec p_\perp$.  We also have that $p^+ = z\, k^+$ and $p^- = (\vec p_\perp^2 + m_d^2)/p^+$, where in our case the daughter mass $m_d = M_t$ if it is a top and zero otherwise.  In this frame, if the initiating particle is a fermion, the fragmentation function to observe particle $j$ emitted from it is
\begin{align}
D_{ij}(z) = \frac 1 z \int d^2 p_\perp \int \frac{dx^- d^2 x_\perp}{2 (2\pi)^3} e^{ik^+x^-/2} \dfrac{1}{2 N_{c,i}}\Tr \sum_{X} \langle 0 | \bnh \chi_i(0,x^-,x_\perp) | j(p)\, X \rangle \langle j(p)\, X | \chi_i^\dagger(0) | 0\rangle \, \Big |_{p^{Xj}_\perp = 0}.
\label{eq:collinsDef}
\end{align}
We average over spin and color of the initiating particle (in our case $N_c$ = 2 for top and higgs, $N_c$ = 1 for the bottom).  Naively, $D_{ij}(z)$ depends on the plus-momentum of both mother ($k^+$) and daughter ($p^+$).  However, we have boost invariance in directions orthogonal to the $\perp$ direction.  Thus, it can only depend on $z(\equiv p^+/k^+)$.

The difficulty with the form of Eq.~\ref{eq:collinsDef} is that it fixes an unobservable quantity, the mother's perp momentum, $k_\perp = 0$.  A better operational definition fixes the axis of the observed daughter's momentum.  Thus, we rotate and boost to a $p_\perp = 0$ frame with $p^+$ unchanged.  The mother perp momentum becomes $\vec k_\perp = -\vec p_\perp/z$.  Doing this gives 
\begin{align}
D_{ij}(z) = z \int \frac{dx^-}{ 4\pi} e^{ik^+x^-/2} \dfrac{1}{2 N_{c,i}}\Tr \sum_{X} \langle 0 | \bnh \chi_i(0,x^-,0_\perp) | j(p)\, X \rangle \langle j(p)\, X | \chi_i^\dagger(0) | 0\rangle \, \Big |_{p^{j}_\perp = 0}.
%\label{eq:collinsDef}
\end{align}
In practice, we compute purely in momentum space and thus calculate the following quantities,
\begin{align}
D^{\text{fermion}}_{ij}(z) &= \frac{z}{Q} \dfrac{1}{2 N_{c,i}}\Tr \int d\Pi_{X} \, \delta \left(1-\frac{p_X^+ + p^+}{Q}\right) \langle 0 | \bnh \chi_i(k) | j(p)\, X \rangle \langle j(p)\, X | \chi_i^\dagger(k) | 0\rangle \, \Big |_{p^{j}_\perp = 0} \nn \\
D^{\text{scalar}}_{ij}(z) &= z \dfrac{1}{N_{c,i}}\Tr \int d\Pi_{X} \, \delta \left(1-\frac{p_X^+ + p^+}{Q}\right) \langle 0 |  \phi_i(k) | j(p)\, X \rangle \langle j(p)\, X | \phi_i^\dagger(k) | 0\rangle \, \Big |_{p^{j}_\perp = 0},
\label{eq:momff}
\end{align}
where $d\Pi_X$ is the phase space of the $X$ particles in the semi-inclusive final state subject to the constraints from the $\delta$-function and on-shellness.

We get a very useful set of constraints from the set of sum rules that massless fragmentation functions satisfy up to order $\alpha_y  \log(\mu^2/m_t^2)$,  which follow from the conservations of probability and momentum:
\begin{align}
	&\frac 1 2 \sum_j \int dz \, D^{R}_{ij}(z) + \int dz \, D^{V}_{ii}(z) = 1 \nn \\
	&\sum_j \int dz \, z\,D^{R}_{ij}(z) + \int dz \, z\, D^{V}_{ii}(z) = 1
\label{eq:ffconserv}
\end{align}
where $R$ stands for fragmentation with real-emission while $V$ indicates self-fragmentation.  Including the effects of finite $M_t$, the probabilistic interpretation of $D_{ij}$ breaks down.  Contributions due to the finite width $\Gamma_t$ give positive contributions to $D_{tb},\, D_{th}$.  However, in our $\overline{\text{MS}}$ scheme, there is no compensating negative term in $D_{tt}$.  Furthermore, the $\mo(\alpha_y)$ corrections (without log enhancement) arising from the ultracollinear contributions contain strictly negative terms ({\it cf}.~Table \ref{tbl:ffs}) allowing $D_{ij}$ to take negative values for some range of $\mu$ and $z$.

%%%%%%%%%%%%%%%%%%%%%%%%%%%%%%%%%%%%%%%%%%%
\section{Soft $\&$ Zero-Bin Contribution}
\label{app:soft}
%%%%%%%%%%%%%%%%%%%%%%%%%%%%%%%%%%%%%%%%%%%

\textit{Throughout the following calculations of this Appendix we note the different notation used to label the daughter and mother momenta with respect to the one used in the definition of the fragmentation functions in \ref{app:ff}.}

The calculation of the soft contribution to a graph can come from the Full Theory calculation, followed by taking the appropriate soft and collinear limits for the graph's momenta.  We recall that it is only in the SCET1 theory that a vertex with a soft Higgs is kinematically allowed and leading-power.  Thus, we can work in the simpler $M_t$ = 0 case.  As an example we look at the bare $b \tor t$ fragmentation function,
\begin{align}
	D_{bt} :
	 \begin{tikzpicture}[baseline=-0.35 ex]
\begin{feynman}
\vertex (a)[crossed dot] {};
\vertex [right=of a] (b);
\vertex [above right=of b] (c1);
\vertex [ right=of c1] (c2);
\vertex [below right=of c2] (d);
\vertex [ right=of d] (e)[crossed dot]{};
\diagram* {
    (a) -- [with arrow=0.5,arrow size=0.15em,momentum=\(p+k\)] (b) -- [double, double distance=0.5ex, thick, momentum=\(k\)] (c1),
    (c2) -- [double, double distance=0.5ex, thick, momentum=\(k\)] (d) -- [with arrow=0.5,arrow size=0.15em,momentum=\(p+k\)] (e),
    (b) -- [dashed, half right, looseness=1, insertion={[solid]0.5}, momentum'=\(p\)] (d)
};
\end{feynman}
\end{tikzpicture}
\end{align}
\begin{align}
D_{0,bt}=\dfrac{z}{2 Q} \frac{y_t^2}{(2\pi)^3} \int \dfrac{d^2 p_\perp dp^+}{2p^+}&\Tr{\bnh\s p \s k \s p} \dfrac{\dfun{1 - (p^+ - k^+)/Q}}{\left((p+k)^2\right)^2 },
\end{align}
with $k^- = 0$, $k_\perp = 0$, we compute the trace:
\begin{align}
D_{0,bt}=\dfrac{z}{2 Q} \frac{y_t^2}{(2\pi)^3} \int \dfrac{d^2 p_\perp dp^+}{2p^+}\,\dfun{1 - (p^+ - k^+)/Q}& \dfrac{p^+ k^+ p^-}{2(k^+ p^-)^2 }.
\end{align}
We now take $p$ to have the scaling of a soft momentum, $(p^+,p^-,p_\perp)\sim (\lambda^2,\lambda^2,\lambda^2)$, where $\lambda$ is the SCET bookkeeping power counting parameter.  We can also replace $p^- = \vec p_\perp^{\,2}/p^+$ by the Higgs on-shellness and use the fact that $k = z\,Q$:
\begin{align}
D_{0,bt}^\text{soft}=\dfrac{1}{8 Q^2} \frac{y_t^2}{(2\pi)^3} \dfun{1 - z} \int d^2 p_\perp dp^+\,& \dfrac{p^+}{\vec p_\perp^{\,2}},
\label{eq:soft}
\end{align}
which thus gives the soft contribution to $D_{0,bt}$.

For the zero-bin contribution of the same graph, we compute it in the collinear sector of SCET1,
\begin{align}
D_{0,bh}^\text{zero}=\dfrac{z}{2 Q} \frac{y_t^2}{(2\pi)^3} \int \dfrac{d^2 p_\perp dp^+}{2p^+}& \text{Tr}\left\lbrace\bnh\left(p^+ + k^+\right)\nh\left(\dfrac{-\s{k}_\perp}{k^+}+\dfrac{\s{p}_\perp+\s{k}_\perp}{p^+ + k^+}\right)\bnh\nh k^+\right. \nonumber \\	&\qquad \left.\bnh\left(\dfrac{-\s{k}_\perp}{k^+}+\dfrac{\s{p}_\perp+\s{k}_\perp}{p^+ + k^+}\right) \left(p^+ + k^+\right)\nh\right\rbrace \dfrac{\dfun{1 - (p^+ - k^+)/Q}}{\left((p+k)^2 \right)^2 },
\end{align}
and then take the soft limit for $p$.  We also impose on-shellness for $p$ and that  $k_\perp=0$, $k^-=0$,
\begin{align}
D_{0,bh}^\text{zero}=\dfrac{z}{2 Q} \frac{y_t^2}{(2\pi)^3} \dfun{1 - z}\int \dfrac{d^2 p_\perp dp^+}{2p^+}& k^+ \vec p_\perp^{\,2}\Tr{\left(\frac{\s{\bar n} \s{n}}{4} \right)^3} \dfrac{1}{4\left(k^+p^- \right)^2 }.
\end{align}
Computing the trace:
\begin{align}
D_{0,bh}^\text{zero}=\dfrac{1}{8 Q^2} \frac{y_t^2}{(2\pi)^3} \dfun{1 - z} \int d^2 p_\perp dp^+\,& \dfrac{p^+}{\vec p_\perp^{\,2}}.
\label{eq:zerobin}
\end{align}
We thus see from equations \ref{eq:soft} and \ref{eq:zerobin}, that $\boxed{D_{0,bh}^\text{zero} \equiv D_{0,bh}^\text{soft}}$. Thus the soft modes do not affect our results as they are cancelled by the zero-bin modes.  The same result holds for the other anomalous dimensions, as well.

%%%%%%%%%%%%%%%%%%%%%%%%%%%
\section{Calculational Details}
\label{app:details}
%%%%%%%%%%%%%%%%%%%%%%%%%%%

\textit{Throughout the following calculations of this Appendix we note the different notation used to label the daughter and mother momenta with respect to the one used in the definition of the fragmentation functions in \ref{app:ff}.}

%%%%%%%%%%%%%%%%%%%%%%%%%%%
\subsection{Calculation of $D_{th}$ in Intermediate Theory (SCET2 with  $M_t\neq 0$)}
The diagram for $D_{th}$ calculation is the following:
\begin{align}
	D_{th} :
	 \begin{tikzpicture}[baseline=-0.35 ex]
\begin{feynman}
\vertex (a)[crossed dot] {};
\vertex [right=of a] (b);
\vertex [above right=of b] (c1);
\vertex [ right=of c1] (c2);
\vertex [below right=of c2] (d);
\vertex [ right=of d] (e)[crossed dot]{};
\diagram* {
(a) -- [double,double distance=0.5ex,thick,with arrow=0.5,arrow size=0.15em,momentum=\(p+k\)](b) -- [scalar, momentum=\(p\)](c1),
(c2) -- [scalar, momentum=\(p\)](d)--[double,double distance=0.5ex,thick,with arrow=0.5,arrow size=0.15em,momentum=\(p+k\)](e),
(b)-- [half right, looseness=1,insertion={[size=10pt]0.5},momentum'=\(k\)](d)
};
\end{feynman}
\end{tikzpicture}
\end{align}
Using the operator definition of the fragmentation function as pointed out in \ref{app:ff}, along with the SCET2 Feynman rules, as presented in Section \ref{sec:frag}, in the complex mass scheme, we get:
\begin{align}
	D_{0,th}=\dfrac{z}{2} y_t^2 \int \dfrac{d^4k}{(2\pi)^4}&\text{Tr}\left\lbrace\bnh\left(p^+ + k^+\right)\nh\left(\dfrac{\s{k}_\perp}{k^+}-\dfrac{\s{p}_\perp+\s{k}_\perp}{p^+ + k^+}-\dfrac{\sqrt{M^2_t  -i M_t \Gamma_t}}{p^+ +k^+}\right)\bnh\nh k^+ \right. \nonumber \\	&\qquad \left.\left(\dfrac{\s{k}_\perp}{k^+}-\dfrac{\s{p}_\perp+\s{k}_\perp}{p^+ + k^+}+\dfrac{\sqrt{M^2_t  +i M_t \Gamma_t}}{p^+ + k^+}\right)\bnh \left(p^+ + k^+\right)\nh\right\rbrace 2\pi \dfun{k^2}\dfrac{\dfun{Q - p^+ - k^+}}{\left((p+k)^2 - M_t^2\right)^2 + M_t^2 \Gamma^2},
\end{align}
where the ``0'' subscript denotes that this is the bare fragmentation function.  We follow the steps outlined in \ref{subsec:otlffc}, performing the calculation in the reference frame, discussed in \ref{app:ff} where daughter's perp momentum is zero ($p_\perp = 0$): 
\begin{align}
	D_{0,th}&=\dfrac{z}{2} y_t^2 \int \dfrac{d^4k}{(2\pi)^4}\nonumber\\
 & \text{Tr}\left\lbrace\bnh\left(p^+ + k^+\right)\nh\left(\dfrac{p^+ \s{k}_\perp}{(p^+ + k^+)k^+} -\dfrac{\sqrt{M^2_t  -i M_t \Gamma_t}}{p^+ +k^+}\right)\bnh\nh k^+\left(\dfrac{p^+ \s{k}_\perp}{(p^+ + k^+)k^+} +\dfrac{\sqrt{M^2_t  +i M_t \Gamma_t}}{p^+ +k^+}\right)\right. \nonumber \\
	&\qquad \left.\bnh \left(p^+ + k^+\right)\nh \right\rbrace 2\pi \dfun{k^2}\dfrac{\dfun{Q - p^+ - k^+}}{\left((p+k)^+(p+k)^- - k_\perp^2  - M_t^2\right)^2 + M_t^2 \Gamma^2}.
\end{align}
The next step is to calculate the trace and simplify the Dirac structure:
\begin{align}
	D_{0,th}=\dfrac{z}{2} y_t^2 \int \dfrac{d^4k}{(2\pi)^3}\left(p^+ + k^+\right)^2 k^+ \left(\dfrac{k_\perp^2 p_+^2 + M_t^2 k_+^2 - M_t \Gamma k_+^2}{k^2_+ (p_+ + k_+)^2}\right)\dfrac{ \dfun{k^2}\dfun{Q - p^+ - k^+}}{\left((p+k)^+(p+k)^- - k_\perp^2  - M_t^2\right)^2 +M_t^2 \Gamma^2}.
\end{align}
We integrate $k^-$ with $k$ on-shell and also $p^- = 0$ due to $p$ on-shellness since we are in the daughter's reference frame:
\begin{align}
D_{0,th}=	\dfrac{z}{2} \dfrac{y_t^2}{2} \int \dfrac{d^2k_\perp dk^+}{(2\pi)^3} \left(k_\perp^2 \dfrac{p_+^2}{k^2_+} + M_t^2\right)\dfrac{ \dfun{Q - p^+ - k^+}}{\left((p+k)^+\dfrac{k_\perp^2}{k^+} - k_\perp^2  - M_t^2\right)^2 +  M_t^2 \Gamma^2} + \mathcal{O}\left(\alpha^2_Y\right).
\end{align}
Here in this step we focused only on the terms that are leading order in $\alpha_y$, any terms of $\mathcal{O}\left(\alpha^2_y\right)$ are subsequently dropped. Defining $z\equiv\dfrac{p^+}{Q}$ and integrating over $k^+$ with the remaining delta function, $\dfun{Q - k^+ - p^+}$:
\begin{align}
	D_{0,th}=\dfrac{z}{2} \dfrac{y_t^2}{2} \int \dfrac{d^2k_\perp}{(2\pi)^3} \left(k_\perp^2 \dfrac{z^2}{(1-z)^2} + M_t^2\right)\dfrac{ \dfrac{(1-z)^2}{z^2}}{\left(k_\perp^2 - M_t^2 \dfrac{1-z}{z}\right)^2+  M_t^2 \Gamma^2  \dfrac{(1-z)^2}{z^2}}.
\end{align}
Regularizing the UV divergence through Dimensional Regularization in $d=2-2\varepsilon$:
\begin{align}
	D_{0,th}=\dfrac{z}{2} \dfrac{y_t^2}{2}\mu^{2\varepsilon} \int \dfrac{d^d k_\perp}{(2\pi)^3} \dfrac{ \left(k_\perp^2 + M_t^2\dfrac{(1-z)^2}{z^2}\right)}{\left(k_\perp^2 - M_t^2 \dfrac{1-z}{z}\right)^2+  M_t^2 \Gamma^2  \dfrac{(1-z)^2}{z^2}},
\label{eq:ucexplicit} 
\end{align}
Integrating $k_\perp$ from $[0,\infty)$ and expanding at $\varepsilon =0$:
\begin{align}
	D_{0,th}= \dfrac{\alpha_y}{4\pi}\left(z \dfrac{1}{\varepsilon} +z \log\left(\dfrac{\mu^2}{M_t^2 \dfrac{1-z}{z} \sqrt{1+\dfrac{\Gamma^2}{M_t^2}}}\right) - \dfrac{M_t}{\Gamma_t}\arctan\left(\dfrac{\Gamma_t}{M_t}\right) +\pi\dfrac{M_t}{\Gamma_t} \right).
\label{eq:baredth}
\end{align}
We note that the $M_t^2$ term in the numerator of Eq.~\ref{eq:ucexplicit} contributes the $z \log[z/(1-z)]$ in $D_{0,th}$ that one can spot in Eq.~\ref{eq:baredth}.  This is precisely the ultracollinear contribution.  It relies on keeping finite $M_t$ in the calculation, but it persists after the fact even if we take $M_t \tor 0$.
The final form of the fragmentation function is given after keeping the LO terms of $\dfrac{\Gamma_t}{M_t}\rightarrow 0$ (to $\mo(\alpha_y)$, $\Gamma_t = \alpha_y M_t/4$) and subtracting the $\dfrac{1}{\varepsilon}$ term. Therefore, we have the renormalized fragmentation function, $D_{th}$:
\begin{align}
	\boxed{D_{th} = \dfrac{\alpha_y}{4\pi}\left(z \log\left(\dfrac{\mu^2}{M_t^2} \right)-z\log\left(\dfrac{1-z}{z}\right) - 1 +\pi\dfrac{M_t}{\Gamma_t} \right)}
\end{align}

%%%%%%%%%%%%%%%%%%%%%%%%%%%%%%%%%
\subsection{Calculation of $D_{hh}$ in Intermediate Theory with  $M_t\neq 0$ }
%%%%%%%%%%%%%%%%%%%%%%%%%%%%%%%%%

The diagram for $D_{hh}$ calculation is the following:
\begin{align}
	 D_{hh}: \begin{tikzpicture}[baseline=-0.35 ex]
\begin{feynman}
\vertex (a)[crossed dot] {};
\vertex [right=of a] (b);
\vertex [right=of b] (d);
\vertex [ right=of d] (e)[crossed dot]{};
\diagram* {
(a) -- [scalar,momentum=\(p\)](b),
(b)-- [fermion,half right, looseness=1,momentum'=\(p-k\)](d),
(b)-- [double,double distance=0.5ex,thick,with arrow=0.5,arrow size=0.15em,half left, looseness=1,momentum=\(k\)](d),
(d)-- [scalar,momentum=\(p\)](e)
};
\end{feynman}
\end{tikzpicture}
\end{align}
We follow the steps outlined in \ref{subsec:otlffc} and we calculate the residue of the expression represented by this graph:
\begin{align}
	\text{Res}\left[\Sigma\left(p\right)\right] =\lim_{p^2\rightarrow 0} \dfrac{d}{dp^2}\left\lbrace \left(p^2\right)^2 \cdot \dfrac{1}{p^2}\Sigma\left(p\right)\cdot \dfrac{1}{p^2}\right\rbrace = \lim_{p^-\rightarrow 0} \dfrac{dp^-}{dp^2}\dfrac{d}{dp^-} \Sigma\left(p\right)=\lim_{p^-\rightarrow 0} \dfrac{1}{p^+}\dfrac{d}{dp^-} \Sigma\left(p\right),
\end{align}
where $ \Sigma\left(p\right)$ is given by:
\begin{align}
 \Sigma\left(p\right)=(-1)y_t^2 \int \dfrac{d^4k}{(2\pi)^4}&\text{Tr}\left\lbrace\left(\dfrac{-\s{k}_\perp}{p^+-k^+}-\dfrac{\s{k}_\perp}{k^+} - \dfrac{M_t}{k^+}\right)\bnh\nh\left(\dfrac{-\s{k}_\perp}{p^+-k^+}-\dfrac{\s{k}_\perp}{k^+} + \dfrac{M_t}{k^+}\right)\bnh\nh\right\rbrace \nonumber\\
 &\dfrac{k^+}{k^+ k^- -k^2_\perp-M_t^2} \left(\dfrac{(p-k)^+}{(p-k)^+ (p-k)^- - k^2_\perp}\right),
\end{align}
and the $(-1)$ comes from the fermion loop factor. The derivative gives:
\begin{align}
 \text{Res}\left[\Sigma\left(p\right)\right]=\lim_{p^-\rightarrow 0} \dfrac{1}{p^+}\dfrac{d}{dp^-} \Sigma\left(p\right) &=-y_t^2 2\int \dfrac{d^4k}{(2\pi)^4}\dfrac{1}{p^+}\left(\dfrac{p_+^2k^2_\perp}{k^+(p^+-k^+)} - \dfrac{M_t^2 (p^+-k^+) }{k^+}\right)\nonumber\\
&\dfrac{(k-p)^+}{k^+ k^- -k^2_\perp-M^2_t} \left(\dfrac{1}{(p-k)^+ (-k^-) - k^2_\perp}\right)^2\nonumber\\
&=-y_t^2 2\int \dfrac{d^4k}{(2\pi)^4}\dfrac{1}{p^+}\left(\dfrac{p_+^2k^2_\perp}{k^+(p^+-k^+)} - \dfrac{M_t^2 (p^+-k^+) }{k^+}\right)\nonumber\\
&\dfrac{1}{k^+ k^- -k^2_\perp-M^2_t}\dfrac{1}{(k^+-p^+)} \left(\dfrac{1}{k^- - \dfrac{k^2_\perp}{(k^+-p^+)}}\right)^2.
\end{align}
Above, the trace is $\text{Tr}\left\lbrace\bnh\nh\bnh\nh\right\rbrace = 2$. Using the residue theorem, the integral over $k^-$ can be calculated by expanding the integrand around the pole, $k^- = (k_\perp^2+M_t^2)/k^+$. The integral is nonzero if $p^+>k^+>0$. Thus,
\begin{align}
	 \text{Res}\left[\Sigma\left(p\right)\right]=&-\dfrac{y_t^2}{2}2\int \dfrac{d^2k_\perp dk^+}{(2\pi)^3}\dfrac{1}{p^+}\left(\dfrac{p_+^2k^2_\perp}{k^+(p^+-k^+)} - \dfrac{M_t^2 (p^+-k^+) }{k^+}\right)\dfrac{1}{(k^+-p^+)} \dfrac{1}{k^+}\left(\dfrac{1}{\dfrac{k^2_\perp}{k^+}+\dfrac{M_t^2}{k^+} - \dfrac{k^2_\perp}{(k^+-p^+)}}\right)^2 \nonumber \\
	&=-\dfrac{y_t^2}{2} 2\int \dfrac{d^2k_\perp dk^+}{(2\pi)^3}\dfrac{1}{p^+}\left(\dfrac{p_+^2k^2_\perp}{k^+(p^+-k^+)} - \dfrac{M_t^2 (p^+-k^+) }{k^+}\right)\dfrac{1}{(k^+-p^+)k^+} \left(\dfrac{1}{\dfrac{M_t^2}{k^+} -\dfrac{k^2_\perp p^+}{k^+(k^+-p^+)}}\right)^2 \nonumber \\
		&=-\dfrac{y_t^2}{2} 2\int \dfrac{d^2k_\perp dk^+}{(2\pi)^3}\dfrac{1}{p^+}\left(-k^2_\perp + M_t^2\dfrac{(k^+-p^+)^2}{(p^+)^2}\right)\left(\dfrac{1}{k^2_\perp-\dfrac{M_t^2 (k^+-p^+)}{p^+} }\right)^2.
\end{align}
The $k_\perp$ in the numerator of the expression above is not Euclidean. Thus we need to rewrite the expression as follows:
\begin{align}
\text{Res}\left[\Sigma\left(p\right)\right]=-\dfrac{y_t^2}{2} 2\int \dfrac{d^2k_\perp dk^+}{(2\pi)^3}\dfrac{1}{p^+}\left(k^2_\perp + M_t^2\dfrac{(k^+-p^+)^2}{(p^+)^2}\right)\left(\dfrac{1}{k^2_\perp-\dfrac{M_t^2 (k^+-p^+)}{p^+} }\right)^2.
\end{align}
Integrating first over $k_\perp$ on $\left[0,\infty \right)$ using Dimensional Regularization:
\begin{align}
& \text{Res}\left[\Sigma\left(p\right)\right] = -\dfrac{y_t^2}{2} 2\int_0^{p^+} \dfrac{dk^+}{8(\pi)^2} \frac{\pi  (k^+-p^+) \mu ^{2 \epsilon } \csc (\pi  \epsilon ) (k^+ \epsilon -2 p^+) \left(\frac{p^+}{M_t^2 (p^+-k^+)}\right)^{\epsilon }}{(p^+)^2 (\epsilon -2)}.
\end{align}
Integrating over $k^+$ and expanding to $\mo(\epsilon^0)$, we get the 1-loop contribution as:
\begin{align}
	 \text{Res}\left[\Sigma\left(p\right)\right] = \dfrac{\alpha_y}{2\pi}\left[ - \log \left(\dfrac{\mu^2}{M_t^2} \right)-\dfrac{1}{\epsilon }-\dfrac{1}{2}\right].
\end{align}
Working toward the fragmentation function, we restore the $z$ dependence, through a $\delta(1-z)$ as we have in the self-fragmentation:
\begin{align}
	 D_{0,hh} = \dfrac{\alpha_y}{2\pi}\left[ - \log \left(\dfrac{\mu^2}{M_t^2} \right)-\dfrac{1}{\epsilon }-\dfrac{1}{2}\right] \delta(1-z).
\end{align}
The final form of the renormalized fragmentation function, after subtracting the $\dfrac{1}{\varepsilon}$, is
\begin{align}
	\boxed{D_{hh} (z)= \dfrac{\alpha_y}{2\pi}\left[ - \log \left(\dfrac{\mu^2}{M_t^2} \right)-\dfrac{1}{2}\right] \delta\left(1-z\right).}
\end{align}
The -1/2 term is the ultracollinear contribution in this case.

\newpage
%%%%%%%%%%%%%%%%%%%%%%%%%%%%%%%%%%%
\section{Mellin Space Anomalous Dimensions}
\label{app:ad}
%%%%%%%%%%%%%%%%%%%%%%%%%%%%%%%%%%%
As we saw in \ref{subsec:rmyukawa}, performing the running requires us to calculate the anomalous dimension matrix in Mellin space. This can be done at LL and NLL-UC, as dictated in equations ~\ref{eq:llad},~\ref{eq:nllucad}, using the matrix elements for fragmentation functions in Mellin space from \ref{eq:dijmellin}. We list the expressions for anomalous dimension matrix elements at LL and NLL-UC in the tables below:\\
\begin{table}[H]
\resizebox{0.3\textwidth}{!}{
 \begin{subtable}{0.3\textwidth}
    \centering
     \begin{tabular}{|c|c|}
    \hline
     & \\
            & $\gamma^{\text{LL}}_{bi}$ for $i \in \left\lbrace h,b,t \right\rbrace$ \\
            & \\
          \hline   
    &  \\
     $\gamma^{\text{LL}}_{bh}$  &  $  \dfrac{\alpha_y}{\pi} \dfrac{(s-1)}{ s^2}  $\\
               &  \\
        \hline
        &  \\
         $\gamma^{\text{LL}}_{bb}$ &  $  -\dfrac{\alpha_y}{\pi} \dfrac{\left(s^3+s^2+2\right)}{2s^2 (s+1)} $\\
        &  \\
         \hline
                    &  \\
        $\gamma^{\text{LL}}_{bt}$ &  $ \dfrac{\alpha_y}{\pi} \dfrac{1}{\left(s^2+s\right)} $\\
        &  \\
        \hline
    \end{tabular}
  \end{subtable}}
  \hfill
  \resizebox{0.3\textwidth}{!}{
 \begin{subtable}{0.3\textwidth}
    \centering
     \begin{tabular}{|c|c|}
    \hline
     & \\
            & $\gamma^{\text{LL}}_{hi}$ for $i \in \left\lbrace h,b,t \right\rbrace$ \\
            & \\
          \hline   
    &  \\
     $\gamma^{\text{LL}}_{hh}$  &  $ -\dfrac{\alpha_y}{\pi}\dfrac{\left(s^2+1\right)}{s^2} $\\
               &  \\
        \hline
        &  \\
         $\gamma^{\text{LL}}_{hb}$ &  $ \dfrac{\alpha_y}{\pi}\dfrac{(s-1)}{ s^2} $\\
        &  \\
         \hline
                    &  \\
        $\gamma^{\text{LL}}_{ht}$ &  $ \dfrac{\alpha_y}{\pi}\dfrac{1}{ s} $\\
        &  \\
        \hline
    \end{tabular}
  \end{subtable}}
  \hfill
   \resizebox{0.3\textwidth}{!}{
 \begin{subtable}{0.3\textwidth}
    \centering
     \begin{tabular}{|c|c|}
    \hline
     & \\
            & $\gamma^{\text{LL}}_{ti}$ for $i \in \left\lbrace h,b,t \right\rbrace$ \\
            & \\
          \hline   
    &  \\
     $\gamma^{\text{LL}}_{th}$  &  $ \dfrac{\alpha_y}{\pi}\dfrac{1+3s}  {4(s^2+s)} $\\
               &  \\
        \hline
        &  \\
         $\gamma^{\text{LL}}_{tb}$ &  $ \dfrac{\alpha_y}{\pi} \dfrac{3+s}{4(s^2+s)} $\\
        &  \\
         \hline
                    &  \\
        $\gamma^{\text{LL}}_{tt}$ &  $ -\dfrac{\alpha_y}{\pi}\dfrac{1}{4} $\\
        &  \\
        \hline
    \end{tabular}
  \end{subtable}}
\caption{Mellin Transforms of the Anomalous Dimension at LL.}
\label{tab:gammall}
\end{table}
  
\begin{table}[H]
 \resizebox{\textwidth}{!}{
    \begin{tabular}{|c|c|}
    \hline
            & \\
            & $\gamma^{\text{NLL-UC}}_{bi}$ for $ i\in \left\lbrace h,b,t \right\rbrace$ \\
    &  \\
    \hline
    &  \\
        $\gamma^{\text{NLL-UC}}_{bh}$  &  $\alpha_y^2\dfrac{\left(-2 (s+1) s^4 H(s+1)-\left(5 s^4+10 s^3+9 s^2+12 s+8\right) s H(s)+2 s^7-5 s^6-2 s^5+35 s^4+38 s^3+26 s^2+32 s+16\right)}{8 \pi ^2 (s+1)^3 s^5}$\\
               &  \\
        \hline
        &  \\
         $\gamma^{\text{NLL-UC}}_{bb}$ &  $ -\alpha^2_y\dfrac{\left(4\left(5 s^4+12 s^3+11 s^2+12 s+8\right) s H(s)+27 s^8+81 s^7+117 s^6+51 s^5-144 s^4-148 s^3-48 s^2-112 s-64\right)}{32 \pi ^2 s^5 (s+1)^3}$\\
        &  \\
         \hline
                    &  \\
        $\gamma^{\text{NLL-UC}}_{bt}$ &  $\alpha^2_y \dfrac{\left(\left(5 s^4+10 s^3+9 s^2+12 s+8\right) s H(s)+\left(9 s^6+13 s^5-24 s^4-38 s^3-22 s^2-32 s-16\right)\right)}{8 \pi ^2 s^4 (s+1)^3} $\\
        &  \\
        \hline
         \end{tabular}}
    \caption{Mellin Transforms of the Anomalous Dimension for $b\rightarrow\left\lbrace h,b,t\right\rbrace$ at NLL-UC.}
    \label{tab:gammanllucb}
\end{table}

\begin{table}[H]
   \resizebox{\textwidth}{!}{
    \begin{tabular}{|c|c|}
    \hline
            & \\
            & $\gamma^{\text{NLL-UC}}_{hi}$ for $i\in \left\lbrace h,b,t \right\rbrace$\\
    &  \\
    \hline
    & \\
    &  \\
       $\gamma^{\text{NLL-UC}}_{hh}$  &  $-\alpha_y^2\dfrac{\left(4 (s+1) s^4 H(s+1)+2 \left(11 s^4+22 s^3+11 s^2+8 s+8\right) s H(s)+11 s^7+22 s^6-11 s^5-78 s^4-80 s^3-44 s^2-48 s-32\right)}{16 \pi ^2 s^5(s+1)^2}$\\
               &  \\
       \hline
               &  \\
        &  \\
        $ \gamma^{\text{NLL-UC}}_{hb}$ & $\alpha^2_y \dfrac{ \left(-\left(11 s^4+24 s^3+13 s^2+8 s+8\right) s H(s)+\left(3 s^6-6 s^5+7 s^4+28 s^3+8 s^2+20 s+16\right)\right)}{8 \pi ^2 s^5 (s+1)^2}  $\\
              &  \\
      \hline
              & \\
        $\gamma^{\text{NLL-UC}}_{ht}$  &  $\alpha^2_y \dfrac{\left(\left(11 s^4+22 s^3+11 s^2+8 s+8\right) s H(s)-\left(11 s^5+37 s^4+44 s^3+18 s^2+24 s+16\right)\right)}{8 \pi ^2 s^4 (s+1)^2} $\\
               &  \\
        \hline
         \end{tabular}
         }
    \caption{Mellin Transforms of the Anomalous Dimension for $h\rightarrow\left\lbrace h,b,t\right\rbrace$ at NLL-UC.}
    \label{tab:gammanlluch}
\end{table}
\begin{table}[H]
     \resizebox{\textwidth}{!}{
    \begin{tabular}{|c|c|}
    \hline
            & \\
            & $\gamma^{\text{NLL-UC}}_{ti}$ for $i\in \left\lbrace h,b,t \right\rbrace$ \\
              &  \\
         \hline
        &  \\
        $\gamma^{\text{NLL-UC}}_{th}$ &  $\alpha^2_y\dfrac{\left(2 s \left(3 s^2+5 s+4\right) H(s)+(s+1) \left(8 s^4 H(s+1)-\left(5 s^4+15 s^3-4 s^2+16 s+16\right)\right)\right)}{16 \pi ^2 s^4 (s+1)^2}$\\
                &  \\
        \hline
               &  \\
         $\gamma^{\text{NLL-UC}}_{tb}$ &  $ \alpha^2_y\dfrac{\left(4 \left(4 s^4+11 s^3+12 s^2+9 s+4\right) s H(s)- \left(13 s^6+49 s^5+59 s^4+39 s^3+68 s^2+92 s+32\right)\right)}{32 \pi ^2 s^4 (s+1)^3} $ \\
                 &  \\
         \hline
                 &  \\
          $\gamma^{\text{NLL-UC}}_{tt}$ &  $-\alpha^2_y\dfrac{\left(s \left(3 s^3+8 s^2+9 s+4\right) H(s)-\left(2 s^4+7 s^3+19 s^2+24 s+8\right)\right)}{8 \pi ^2 s^3 (s+1)^3}$\\
                  &  \\
          \hline
    \end{tabular}
    }
    \caption{Mellin Transforms of the Anomalous Dimension for $t\rightarrow\left\lbrace h,b,t\right\rbrace$ at NLL-UC.}
    \label{tab:gammanlluct}
\end{table}

\newpage
\bibliographystyle{JHEP}
\bibliography{refer}

\end{document}